\documentclass[prd,showpacs,showkeys,nofootinbib,preprint]{revtex4-1}
\usepackage{amsmath,amsfonts,amssymb,color}
\usepackage{bm}% bold math
\usepackage{pgfplots}
\usepackage[colorlinks=true,urlcolor=blue,linkcolor=blue,citecolor=blue]{hyperref}
\usepackage{amsmath,amssymb,color,graphics,graphicx}
\usepackage{bm}% bold math
\usepackage{hyperref} 

\definecolor{dark-green}{rgb}{0,0.7,0}
\definecolor{dark-blue}{rgb}{0,0.2,0.5}
\definecolor{med-blue}{rgb}{0,0.7,1}
\definecolor{mblue}{rgb}{0,0.2,1}
\definecolor{cnc}{rgb}{0.8,0,0}
\definecolor{light-red}{rgb}{1,0.8,0.8}
\definecolor{dark-yellow}{rgb}{1,0.8,0}
\definecolor{light-blue}{rgb}{0.8,0.9,1}
\definecolor{verylight-blue}{rgb}{0.93,0.95,1}
\definecolor{light-yellow}{rgb}{1,0.9,0.8}
\definecolor{grey}{gray}{0.88}

\usepackage{amsfonts}
\begin{document}

%%%%%%%%%%%%%%%%%%%%%%%%%%%%%%%%%%
%%%%%%%%%%%%%%%%%%%%%%%%%%%%%%%%%%
\title{Momentum of light in complex media}
\author{Yuri N. Obukhov}
\affiliation{Russian Academy of Sciences, Nuclear Safety Institute, 
B.Tulskaya 52, 115191 Moscow, Russia
\email{obukhov@ibrae.ac.ru}}

\begin{abstract}
An important issue in phenomenological macroscopic electrodynamics of moving media is the definition of the energy and momentum of the electromagnetic field in matter. Rather surprisingly, this topic has demonstrated a remarkable longevity, and the problem of the electromagnetic energy and momentum in matter remained open, despite numerous theoretical and experimental investigations. We overview the definition of the momentum of light in matter and demonstrate that, for the correct understanding of the problem, one needs to carefully distinguish situations when the material medium is modeled either as a background for light or as a dynamical part of the total system. The status of Minkowski and Abraham energy-momentum tensors of the electromagnetic field is clarified for the two particular types of complex matter, the spinning fluid and the liquid crystal medium, and summarized for the case of general anisotropic moving material media with a linear constitutive law.
% \keywords{classical electrodynamics, energy-momentum tensor, momentum of light, Abraham-Minkowski controversy, moving media, microstructure}
\end{abstract}

\maketitle

\section{Introduction}

Light (electromagnetic wave) carries energy and {\it momentum}. Theoretical aspects of the electromagnetic wave propagation were first developed by Maxwell (1873), who wrote in Sec.~792 of Chap.~20, Vol. 2 of \cite{Maxwell}: ``... in a medium in which waves are propagated there is a pressure in the direction normal to the waves...''. Experimentally this was verified by Lebedev (1901) who measured the pressure of light on a mirror in vacuum \cite{Lebedev}. 

In the modern relativistic framework, the issue of the energy and momentum carried by the electromagnetic wave, and the stress produced by it, amounts to the study of the energy-momentum tensor of the electromagnetic field in a medium \cite{penfield,pla,Lindell1,Lindell2}. More than a century ago, in 1908, Hermann Minkowski developed a relativistically covariant formulation of classical Maxwell-Lorentz electrodynamics and in this framework he gave a solution of the momentum of light problem \cite{minkowski}. However, soon after Minkowski proposed his version of the electromagnetic energy-momentum tensor, Max Abraham entered the dispute with a different expression \cite{abraham1,abraham2}. The main formal difference between the Minkowski and the Abraham energy-momentum tensors is that the latter is symmetric while the former is not. 
 
Electromagnetic field momentum in vacuum has a well known form
\begin{equation}
\bm{p} = {\bm D}\times {\bm B} = \varepsilon_0\,{\bm E}\times {\bm B}.\label{p0}
\end{equation}
The question is: how the momentum of light looks inside a material medium that can have complicated electric and magnetic properties and, in general, can move in an arbitrary way? At the first sight, this seems to be a strange question: after all, physics is an experimental science, and observations should tell us the answer -- so, why there is a problem? However, a puzzling disagreement of the theoretical predictions with observations exists in the electromagnetics of continua, giving rise to what is known a Minkowski-Abraham controversy. 

Most of electromagnetic wave experiments were done for the simple media which are spatially isotropic, homogeneous and static \cite{Milonni,Ramos}. In this case, the properties of matter are characterized by the permittivity $\varepsilon$ and permeability $\mu$ which are combined into a refractive index $n = \sqrt{\varepsilon\mu}$ of the medium. For the Minkowski energy-momentum, the light carries a greater momentum than in vacuum (\ref{p0}), namely, $n{\bm p}$, whereas according to Abraham, the light momentum ${\bm p}/n$ is smaller than in vacuum. The light pressure experiments that extend the classic measurements of Lebedev from vacuum to dielectric media \cite{jones1,jones2,gibson} support Minkowski's answer, whereas the force measurement experiments in moving simple media \cite{james,wlw,ww} seem to support Abraham's solution. 

%\baselineskip=10pt
%\begin{footnotesize}
\begin{table}
\caption{\label{tab_symbols1}Directory of symbols.}
\begin{tabular}{lp{11.5cm}}
\hline\hline
Symbol & Explanation\\
\hline
\hline
\multicolumn{2}{l}{{General quantities (geometry, kinematics)}}\\
\hline
\vspace{-3mm}$x^i$, $(t, \bm{x})$ & Spacetime coordinates, (time and space)\\
\vspace{-3mm}$g_{ij}$ & Spacetime metric\\
\vspace{-3mm}$\delta^i_j$ & Kronecker symbol \\
\vspace{-3mm}$\eta_{ijkl}$ & Totally antisymmetric Levi-Civita tensor \\
\vspace{-3mm}$\epsilon_{ijk}$ & Transversal projection of Levi-Civita tensor \\
\vspace{-3mm}$\ell_{ij}$ & Lorentz group generators \\
\vspace{-3mm}$P_i^j$, $\pi_i^j$ & Projectors\\
\vspace{-3mm}$u^i$ & 4-velocity \\
\vspace{-3mm}$a^i$ & 4-acceleration \\
\vspace{-3mm}$h_{ij}$ & Shear tensor \\
$L^{\rm e}, L^{\rm m}$ & Lagrangians of electromagnetic field and matter\\
\hline
\multicolumn{2}{l}{{Electromagnetic quantities}}\\
\hline
\vspace{-3mm}$J^i$, $(\rho_{\rm e}, \bm{J})$ & Electric 4-current density, (charge density, 3-current density)\\
\vspace{-3mm}$F_{ij}$, $H^{ij}$ & Electromagnetic field strength and excitation tensors\\
\vspace{-3mm}$\bm{E}, \bm{B}$, $\bm{D}, \bm{H}$, ${\mathcal E}_i$, ${\mathcal B}^i$, ${\mathcal H}_i$, ${\mathcal D}^i$ & Electric and magnetic fields and excitations\\
\vspace{-3mm}$\varepsilon_0, \mu_0$ & Electric, magnetic vacuum constants \\
\vspace{-3mm}$c$ & Velocity of light in vacuum\\
\vspace{-3mm}$n$ & Refractive index\\
\vspace{-3mm}$\varepsilon, \mu$ & Relative permittivity and permeability of matter\\
\vspace{-3mm}$\chi^{ijkl}$ & Constitutive tensor \\
\vspace{-3mm}$\varepsilon^{ij}$, $\mu^{-1}_{ij}$, $\gamma_i{}^j$ & Permeability, permittivity, magnetoelectric moduli tensors \\
\vspace{-3mm}$\bm{p}$ & Electromagnetic field momentum density\\
\vspace{-3mm}$\bm{s}$ & Electromagnetic energy flux density (Poynting vector)\\
\vspace{-3mm}$S_a{}^b$ & Electromagnetic (Maxwell) stress tensor\\
${\stackrel {\rm M}\Sigma}{}_k{}^i$, ${\stackrel {\rm A}\Sigma}{}_k{}^i$  & Minkowski, Abraham energy-momentum tensors\\
\hline
\multicolumn{2}{l}{{Material quantities}}\\
\hline
\vspace{-3mm}$\rho$ & Internal energy density of matter\\
\vspace{-3mm}$\nu$ & Number particle density of matter\\
\vspace{-3mm}$s$, $T$ & Entropy, temperature \\
\vspace{-3mm}$X$ & Lin (particle identity) variable\\
\vspace{-3mm}$p$ & Pressure\\
\vspace{-3mm}$b^i_A$, $e^i_\alpha$ & Cosserat triad of matter medium, material vierbein (tetrad) \\
\vspace{-3mm}$\Omega^\alpha{}_\beta$ & Generalized acceleration tensor\\
\vspace{-3mm}$\mu^{AB}$, ${\mathcal S}_{ij}$ & Specific spin density, relativistic spin density tensor\\
\vspace{-3mm}$N^i$ & Director 4-vector\\
\vspace{-3mm}$K_1, K_2, K_3$, ${\mathcal J}$ & Frank's elastic constants, element's moment of inertia\\
\vspace{-3mm}${\mathcal V}$ & Elastic energy density \\
\vspace{-3mm}$\lambda_0,\dots \lambda_5, \lambda^{AB}, \lambda^A_i$ & Lagrange multipliers\\
\vspace{-3mm}$\Psi^A$ & Internal degrees of freedom of matter \\
\vspace{-3mm}$\zeta$ & Magnetic dipole coupling constant \\
\vspace{-3mm}$S_{ij}{}^k$ & Canonical spin tensor of matter\\
$\sigma_k{}^i$, $\kappa_k{}^i$, ${\stackrel {\rm BR}\sigma}{}_k{}^i$ & Canonical, kinetic, symmetrized energy-momentum tensors of matter\\
\hline\hline 
\end{tabular}
\end{table}
%\end{footnotesize}
  
The early theoretical \cite{pauli,wgordon,tamm1939,vonlaue,moller1,balazs,jpgordon,israel,brevik,schmutzer1,schmutzer2} and experimental \cite{jones1,james,ashkin,wlw,ww,jones2,gibson,kristensen} attempts to resolve the controversy were focused on the question which of the two expressions ($n{\bm p}$ vs. ${\bm p}/n$) was the correct one to describe the momentum of light in a homogeneous medium with the refractive index $n$. That resulted in many inconsistent, contradictory, and often confusing statements, before an essential progress was achieved \cite{penfieldhaus1,penfield} by treating the matter and the field as the two coupled subsystems and by making use of the variational approach to derive the total energy-momentum tensor for the resulting closed system. Eventually it became clear that both the Abraham and Minkowski tensors actually give the correct physical results, which however depend on the way how the total energy-momentum tensor of the coupled system is decomposed into ``matter'' and ``field'' pieces \cite{robinson,mikura,maugin,kranys}. Lately, the discussion of the optical momentum in media has been revisited in view of an increasing interest in the study of optical forces in nanotechnology \cite{ashkintweesers,revolutiontweesers,binding} and metamaterials \cite{cloak0,cloak,Kemp:2016}. New theoretical research \cite{loudon1,pla,padgettbarnettloudon,loudon,garrison,milonniboyd2,leonhardt,hindsbarnett,mansuripur5,mansuripur,crenshaw,bradshawboydmilonni,shevchenko2,brevik2010,brevik2012,kempletter,nester1,nester2,Kinsler} and also new experiments \cite{campbell,she,rikken,rikken2,Kundu,Zhang:2015,Zhang:2012,Astrath:2022} have been reported recently, still trying to judge between the Abraham or Minkowski tensors unequivocally. In most cases, the electromagnetic field was discussed alone without analyzing the dynamics of matter, with just a few exceptions \cite{pfeifer,obukhov1,sheppard:2014,sheppard:2016a,sheppard:2016b,Partanen:2019}, where a coupled system ``matter+field'' was studied in the spirit of \cite{penfieldhaus1,penfield}. Most recently a careful reanalysis of the Abraham-Minkowski controversy \cite{barnett1,barnett2,obukhov1,JOPT,Brevik:2017,Brevik:2018a,Brevik:2018b,Brevik:2019} demonstrated that both the Minkowski and Abraham momenta can be in fact measured (hence both are correct), however, under different physical assumptions. One can consistently identify the Abraham momentum as the ``kinetic'' momentum of light in matter, which is most clearly manifest in the Einstein box thought experiment \cite{balazs,Ramos}, whereas the Minkowski momentum is naturally identified as the ``canonical'' momentum being related to the symmetry generator of translations in the Noether sense. Eventually, such a ``final'' resolution of the Abraham-Minkowski controversy has become generally accepted in the modern research \cite{Milonni,Ramos,saldanha,saldanha2,dodin,crenshaw2013,crenshaw2014,webb,kempresolution,baxter,griffiths}. For the recent reviews and more literature on the subject, we can recommend \cite{pfeifer,obukhov1,baxter,griffiths,JOPT,Anghinoni:2022,Kemp:2015,Brevik:2017,Brevik:2018b}.

In the present paper, we give an overview of the subtleties of the Abraham-Minkowski controversy for the case of the moving {\it complex media}, i.e., the matter with microstructure. After briefly reminding in Sec.~\ref{Maxwell} of the relativistic formulation of Maxwell's electrodynamics, the Abraham and Minkowski energy-momentum tensors are explicitly introduced in Sec.~\ref{Two} for complex media with a general linear constitutive relation. The classical electromagnetism theory is recast in Sec.~\ref{Pro} into a relativistic 4-vector formalism with the help of a projector technique. In order to understand more clearly how the Abraham and Minkowski tensors are defined for the complex media, we consider the two models of matter in greater detail. Namely, in Sec.~\ref{weyss} we deal with the spinning fluid, and in Sec.~\ref{crystal} we discuss the liquid crystal medium. The general closed system of the moving complex matter with microstructure coupled to the electromagnetic field with an arbitrary constitutive tensor is analysed in Sec.~\ref{general}, where the resolution of the controversy is explained. The outlook and conclusions are given in Sec.~\ref{conc}.

Our basic notations and conventions follow \cite{obukhov1,Birkbook}. In particular, the indices from the middle of the Latin alphabet $i,j,k,\ldots = 0,1,2,3$ label the 4-dimensional spacetime components, the Greek alphabet is used for anholonomic frame indices $\alpha, \beta, \ldots = 0,1,2,3$, whereas the Latin indices from the beginning of the alphabet $a,b,c,\ldots = 1,2,3$ refer to the 3-dimensional spatial objects and operations (the 3-vectors are also displayed in boldface). The Minkowski metric is defined as $g_{ij}:={\rm diag}(c^2,-1,-1,-1)$. In the 4-dimensional framework spatial components of tensor are raised or lowered by $g_{ab}=-\delta_{ab}$, however, when dealing only with 3-dimensional tensors, one should use the Euclidean 3-metric $\delta_{ab}$ to move the spatial indices. A directory of symbols used in the text can be found in Table~\ref{tab_symbols1}.

\section{Classical electrodynamics}\label{Maxwell}

Maxwell's equations in 3-dimensional notation read \cite{Birkbook}:
\begin{alignat}{2}
\bm{\nabla}\cdot\bm{D} &= \rho_{\rm e}, \qquad
\bm{\nabla}\times\bm{H} - {\frac {\partial\bm{D}}{\partial t}} &= {\bm J},\label{Max2}\\
\bm{\nabla}\cdot\bm{B} &= 0, \qquad
\bm{\nabla}\times\bm{E} + {\frac {\partial\bm{B}}{\partial t}} &= 0.\label{Max3}
\end{alignat}
Here $({\bm D}, {\bm H})$ are the electric and magnetic excitations, and $({\bm E}, {\bm B})$ are the electric and magnetic fields; $(\rho_{\rm e}, {\bm J})$ are the charge and current densities of the sources. 

For {\it simple media} (isotropic matter without microstructure at rest) the excitations and fields are linked by the constitutive relation
\begin{equation}\label{CRsimple}
{\bm D} = \varepsilon\varepsilon_0{\bm E}\quad {\rm and}\quad {\bm H} = {\frac 1{\mu\mu_0}}{\bm B},
\end{equation}
where $\varepsilon$ and $\mu$ are (relative) permittivity and permeability of matter and $\varepsilon_0$ and $\mu_0$ are electric and magnetic vacuum constants. 

A general linear and local constitutive relation for {\it complex media} ($a,b,\dots=1,2,3$) is more nontrivial:
\begin{align}
{D}^a &= \varepsilon_0{\varepsilon^{{ab}}}\,E_b + \gamma_b{}^a\,{B}^b\,,\label{CR1}\\ 
{H}_a &= \mu^{-1}_0{\mu_{ab}^{-1}} {B}^b - \gamma_a{}^b\,E_b.\label{CR2}
\end{align}
The electric and magnetic properties of matter are encoded in 6 relative permittivities $\varepsilon^{ab}= \varepsilon^{ba}$, 6 relative permeabilities $\mu_{ab}=\mu_{ba}$, and 9 magnetoelectric moduli $\gamma_b{}^a$, cf. \cite{Birkbook}.

To discuss the energy-momentum problem for the case of an arbitrary complex medium, it is more convenient to recast Maxwell's theory into an explicitly 4-dimensional covariant form. 

In the relativistic formulation, the electric and magnetic fields $({\bm E}, {\bm B})$ and  the electric and magnetic excitations $({\bm D}, {\bm H})$ form the two 4-dimensional tensors 
\begin{eqnarray}
F_{ij} &=& \left(\begin{array}{cccc}
\!\!0\!\!& \!\!-E_1 \!\!& \!\!-E_2 \!\!& \!\!-E_3 \!\! \\
\!\!E_1 \!\!&\!\!0 \!\!& \!\!B^3 \!\!& \!\!-B^2\!\! \\  
\!\!E_2\!\!& \!\!-B^3\!\!&\!\!0\!\!&\!\!B^1 \!\! \\ 
\!\!E_3\!\!& \!\!B^2\!\!&\!\!-B^1\!\!&\!\!0 \!\!\end{array}\right),\\
H^{ij} &=& \left(\begin{array}{cccc}
\!\!0 \!\!&\!\!D^1 \!\!&\!\!D^2 \!\!&\!\!D^3 \!\! \\
\!\!-D^1\!\!&\!\!0 \!\!&\!\!H_3\!\!&\!\!-H_2 \!\! \\  
\!\!-D^2\!\!&\!\!-H_3\!\!& \!\!0 \!\!&\!\!H_1\!\! \\  
\!\!-D^3\!\!&\!\!H_2\!\!&\!\!-H_1\!\!&\!\!0 \!\!\end{array}\right).
\end{eqnarray}
With the 4-dimensional indices $i,j,k = 0, 1, 2, 3$, we introduce the 4-current vector $J^i = (\rho_{\rm e}, {\bm J})$, and rewrite Maxwell's equations (\ref{Max2}) and (\ref{Max3}) as
\begin{equation}
\partial_jH^{ij}= J^i\,,\quad \partial_iF_{jk} + \partial_j F_{ki} + \partial_k F_{ij}=0.\label{Max4}
\end{equation}

Then the general constitutive relation for local, linear, moving media (\ref{CR1})-(\ref{CR2}) is recast into
\begin{equation}
H^{ij} = {\frac 12}\,\chi^{ijkl}F_{kl}.\label{CR}
\end{equation}
The components of the constitutive tensor encompass all the moduli, $\chi^{ijkl} = \{\varepsilon^{ab}, \mu_{ab},\gamma_b{}^a\}$, and describe all the electric and magnetic properties of matter. The linear constitutive relation (\ref{CR}) arises from the definition of the electromagnetic excitation tensor
\begin{equation}
H^{ij} = -\,2{\frac {\partial L^{\rm e}}{\partial F_{ij}}}\label{Hij}
\end{equation}
for the Lagrangian
\begin{equation}
L^{\rm e} = -\,{\frac 18}\,\chi^{ijkl}\,F_{ij}F_{kl}.\label{Le}
\end{equation}

\section{A Tale of Two Tensors}\label{Two}

In the relativistic framework, the momentum of light ${\bm p}$ is a piece of the energy-momentum tensor. In vacuum, the latter reads:
\begin{equation}
\Sigma_j{}^i = {\frac 1{\mu_0}}\left(- F_{jk}F^{ik} + {\frac 14}\delta_j^iF_{kl}F^{kl}\right).
\end{equation}
The properties of this tensor are well known: it is conserved (without sources), and is symmetric
\begin{equation}
\partial_i\Sigma_j{}^i = 0,\qquad \Sigma_{ij} = \Sigma_{ji}. 
\end{equation}
One recovers (\ref{p0}) as $p_a = -\,\Sigma_a{}^0$.

Before we address the issue of the form of the electromagnetic energy-momentum in a medium, it is instructive to recall the mechanics. The momentum of a point particle reads ${\bm p} = m{\bm v}$. However, for complex systems the ``kinetic'' intuition may be misleading. For example, the momentum of particle in a noninertial reference frame (rotating with an angular velocity $\bm{\omega}$) reads ${\bm p} = m{\bm v} + m{\bm \omega} \times{\bm r}$, whereas for the charged particle moving in an electromagnetic field $\bm{A}$ one has ${\bm p} = m{\bm v} + q{\bm A}$. These expressions can be easily understood if one consistently uses the {\it first principles} -- the Lagrangian approach and the Noether formalism for the symmetries of the theory. 

Let us now turn to the discussion of the status of Minkowski and Abraham energy-momentum tensors and eventually formulate the solution of the Minkowski-Abraham controversy. The crucial point is to distinguish {\it open} and {\it closed} systems.

The physical explanation of the fact that the momentum of light in matter differs from that in vacuum is clear: this happens because the electromagnetic field interacts with a polarizable/magnetiz\-able medium. Depending on the physical conditions in actual experiment, there are two situations when the dynamics of the electromagnetic field in matter is described either as an open or as a closed system. When the medium is fixed externally and treated as a background, the electromagnetic (wave) field represents an {\it open} physical system. When both the medium and the light are dynamical, they together form a {\it closed} system. 

The mechanics of material continua is then an important tool for understanding of the momentum (energy-momentum) of light in a medium, with the help of which one can construct the relativistic energy-momentum tensor $\sigma_k{}^i$ of matter explicitly. For example, for non-interacting particles one finds $\sigma_k{}^i = u^i{\cal P}_k$, with the average 4-velocity of the fluid $u^i$ and the 4-momentum ${\cal P}_k = {\frac \rho{c^2}} u_k$. Taking the collisions into account, this is generalized to $\sigma_k{}^i = u^i{\cal P}_k - (\delta_k^i - \frac{1}{c^2}u_ku^i)p$, with the pressure $p$. If, moreover, the elastic properties are included, one finds $\sigma_k{}^i = u^i{\cal P}_k - (\delta_k^i - \frac{1}{c^2}u_ku^i)p + {\stackrel {\rm a} \sigma}{}_k{}^i$, where the elastic stress tensor ${\stackrel {\rm a} \sigma}{}_k{}^i$ is not symmetric, in general (mind the {\it spin}!).

\subsection{Minkowski tensor}

The Minkowski energy-momentum tensor reads
\begin{equation}
{\stackrel {\rm M} \Sigma}{}_k{}^i = -\,H^{ij}F_{kj} + {\frac 14}H^{jl}F_{jl}\,\delta_k^i.\label{TM}
\end{equation}
Its components are: the energy density $U = {\stackrel {\rm M} \Sigma}{}_0{}^0$
\begin{equation}\label{UM}
U = {\frac 1 2}\left({\bm E}\cdot{\bm D} + {\bm B}\cdot{\bm H}\right),
\end{equation}
the energy flux density (the Poynting vector) $s^a = {\stackrel {\rm M} \Sigma}{}_0{}^a$
\begin{equation}
 {\bm s} = {\bm E}\times{\bm H},\label{sM}
\end{equation}
the electromagnetic field momentum density $p_a = - {\stackrel {\rm M} \Sigma}{}_a{}^0$
 \begin{equation}
{\bm p} = {\bm D}\times{\bm B},\label{pM}
\end{equation}
and the stress tensor $S_a{}^b = {\stackrel {\rm M} \Sigma}{}_a{}^b$
\begin{equation}
S_a{}^b = E_aD^b + H_aB^b - {\frac 1 2}\delta_a^b
\left({\bm E}\cdot{\bm D} + {\bm B}\cdot{\bm H}\right).\label{MM}
\end{equation}
 
The properties of Minkowski tensor are as follows. It is defined for a medium with an {\it arbitrary} constitutive relation (\ref{CR}) from first principles: this is the {\it canonical} Noether current corresponding to the invariance of the system under the spacetime translations. Maxwell's equations (\ref{Max4}) yield the balance law
\begin{equation}\label{BM}
\partial_i\,{\stackrel {\rm M} \Sigma}{}_k{}^i = F_{ki}J^i + {\cal F}_i^{\rm m},
\qquad {\cal F}_i^{\rm m} = {\frac 18}\,F_{mn}F_{pq}\,\partial_k\chi^{mnpq}.
\end{equation}
The Minkowski energy-momentum is not symmetric: the field momentum ${\bm p}$ is not equal to ${\bm s}/c^2$ the energy flux. In other words, there is a nontrivial torque on the right-hand side of the angular momentum balance law,
\begin{equation}\label{nonM}
{\stackrel {\rm M} \Sigma}{}_{[ij]} = {\cal T}_{ij}^{\rm m},
\qquad {\cal T}_{ij}^{\rm m} = {\frac 18}\,F_{mn}F_{pq}\,(\ell_{ij})^{mnpq}_{m'n'p'q'}\chi^{m'n'p'q'},
\end{equation}
where $(\ell_{ij})$ are the Lorentz group generators:
\begin{equation}
(\ell^j{}_k){}^{mnpq}_{m'n'p'q'} = \delta^{[j}_{m'}\delta^m_{k]}
\delta^n_{n'}\delta^p_{p'}\delta^q_{q'} + \delta^m_{m'}\delta^{[j}_{n'}
\delta^n_{k]}\delta^p_{p'}\delta^q_{q'} + \delta^m_{m'}\delta^n_{n'}
\delta^{[j}_{p'}\delta^p_{k]}\delta^q_{q'} + \delta^m_{m'}\delta^n_{n'}
\delta^p_{p'}\delta^{[j}_{q'}\delta^q_{k]}.
\end{equation}

We thus have an {\it open} physical system in which the dynamical electromagnetic field $F_{ij}$ interacts with the ``external'' fixed background encoded in the constitutive tensor $\chi^{ijkl}$. The external force ${\cal F}_i^{\rm m}$ and the torque ${\cal T}_{ij}^{\rm m}$ are nontrivial for an inhomogeneous and anisotropic background. It is worthwhile to notice the traces of the Noether symmetry which is manifest in the presence of the generators of translations $\partial_i$ and rotations $(\ell_{ij})$  in the right-hand sides of (\ref{BM}) and (\ref{nonM}).

\subsection{Abraham tensor}

Given the 4-velocity $u^i$ of a medium (normalized as $u_iu^i = c^2$), the Abraham energy-momentum tensor \cite{obukhov1} is defined by
\begin{eqnarray}
{\stackrel {\rm A} \Sigma}{}_k{}^i &=& -\,{\frac 12}(H^{ij}F_{kj} + F^{ij}H_{kj})
+ {\frac 14}H^{jl}F_{jl}\,\delta_k^i \nonumber\\ &&
+\,{\frac 1{2c^2}}\!\left[\!u^iu_l\!\left(\!F_{jk}H^{jl} - H_{jk}F^{jl}\!\right)
+ \,u_ku^l\!\left(\!F^{ji}H_{jl} - H^{ji}F_{jl}\!\right)\!\right].\label{TA} 
\end{eqnarray}
When the medium is at rest ($u^i = \delta^i_0$): we find same energy density
\begin{equation}
U = {\frac 1 2}\left({\bm E}\cdot{\bm D} + {\bm B}\cdot{\bm H}\right),\label{UA}
\end{equation}
whereas the field momentum and the energy flux satisfy
\begin{equation}
{\bm s} = {\bm E}\times{\bm H},\qquad {\bm p} = {\frac {\bm s}{c^2}} 
= {\frac {{\bm E}\times{\bm H}}{c^2}}.\label{sA}
\end{equation}
The stress tensor reads 
\begin{eqnarray}
S_a{}^b = {\frac 1 2}\left(E_aD^b + E^bD_a + H_aB^b + H^bB_a\right) 
-\,{\frac 12}\delta_a^b\left({\bm E}\cdot{\bm D} +{\bm B}\cdot{\bm H}\right).\label{AA}
\end{eqnarray}
 
The properties of Abraham tensor are as follows. It is defined for a given velocity vector field $u^i$ in an {\it ad hoc} way, not from first principles. However, unlike the Minkowski tensor, the Abraham energy-momentum is symmetric {\it by construction}. The Abraham tensor is not conserved (for any medium):
\begin{align}
\partial_j{\stackrel {\rm A} \Sigma}{}_i{}^{j}+{\cal F}_i^A
+{\cal F}_i^{\rm J}+{\cal F}_i^{\rm m}=0.\label{BA}
\end{align}
Besides the Lorentz ${\cal F}_i^{\rm J} = F_{ij}J^j$ and material ${\cal F}_i^{\rm m}$ forces (\ref{BM}), the {\it Abraham force} ${\cal F}_i^A$ is present in the balance law (\ref{BA}). For the medium at rest, one finds for this additional force, ${\stackrel \circ{\cal F}}_{i}{}^A=(0,-{\stackrel \circ f}{}^A_a)$: 
\begin{align}
\bm{{\stackrel \circ f}}{}^A = \frac{\partial}{\partial t}\left(\bm{{\stackrel {\circ}D}}
\times\bm{{\stackrel {\circ}B}} - \frac{1}{c^2}\bm{{\stackrel {\circ}E}}\times\bm{{\stackrel {\circ}H}}\right) 
+ \frac{1}{2}\bm{\nabla}\times\left(\bm{{\stackrel {\circ}D}}\times\bm{{\stackrel {\circ}E}}
+ \bm{{\stackrel {\circ}B}}\times\bm{{\stackrel {\circ}H}}\right).\label{fA1}
\end{align} 
For an isotropic homogeneous {\it simple medium} it reduces to the famous expression
\begin{equation}\label{fA2}
\bm{{\stackrel \circ f}}{}^A = {\frac {\varepsilon\mu - 1}{c^2}}{\frac
{\partial}{\partial t}}\bm{{\stackrel {\circ}E}}\times\bm{{\stackrel {\circ}H}}.
\end{equation}

\section{Projector formalism in electrodynamics of moving media}\label{Pro}

The above discussion was in the laboratory reference system. In a continuous medium moving with an average 4-velocity $u^i$, it is reasonable to consider another reference system of an observer comoving with the matter. Technically, this amounts to projecting all physical objects to the rest frame of that observer with the help of the projector 
\begin{align}
P^i_j := \delta_j^i - {\frac 1{c^2}}\,u^iu_j.\label{pro}
\end{align}
One can verify that this is an idempotent tensor,
\begin{equation}
P^i_jP^j_k = P^i_k,\label{PP}
\end{equation}
and hence it is a projector on the local three-dimensional hyperplane, orthogonal to the 4-velocity. Since the latter is a timelike vector field, the resulting 3-plane is spacelike. In order to deal with the tensorial objects on this hyperplane, we will need the ``three-dimensional'' projected version
\begin{equation}
\epsilon_{ijk} := {\frac 1c}\,\eta_{ijkl}u^l\label{eps3}
\end{equation}
of the four-dimensional totally antisymmetric Levi-Civita tensor $\eta_{ijkl}$. Its only nontrivial component is $\eta_{0123} = c$. It is worthwhile to notice the useful properties
\begin{align}
\epsilon^{ijk}\epsilon_{lmn} = (P^i_mP^j_l - P^i_lP^j_m)P^k_n + (P^i_nP^j_m - P^i_mP^j_n)P^k_l 
+ (P^i_lP^j_n - P^i_nP^j_l)P^k_m,\label{epep0}\\
\epsilon^{ijn}\epsilon_{kln} = P^i_lP^j_k - P^i_kP^j_l,\qquad \epsilon^{imn}\epsilon_{jmn} = -\,2P^i_j,\qquad
\epsilon^{ijk}\epsilon_{ijk} = -\,6.\label{epep}
\end{align}
By definition, $\epsilon_{ijk}u^k = 0$. This object plays an important role, allowing to introduce a ``cross product'' for the spatial vectors. Namely, given the two spatial vectors $X^i$ and $Y^i$ (such that $X^iu_i = 0$ and $Y^iu_i = 0$), we define a cross product by
\begin{equation}
(X\times Y)_i := \epsilon_{ijk}X^jY^k.\label{cross}
\end{equation}
Obviously, the resulting vector is also a spatial one. Furthermore, for any $Z_i$ we define a {\it generalized curl} by
\begin{equation}
({\rm curl}\,Z)^i := \epsilon^{ijk}\partial_jZ_k,\label{curl}
\end{equation}
which is also a spatial object. On the other hand, a derivative along the timelike 4-velocity is a proper generalization of the {\it time derivative}
\begin{equation}
\dot{Z}^i := u^k\partial_kZ^i.\label{dot}
\end{equation}
A specific example is the 4-acceleration $a^i = \dot{u}^i$. 

For a material continuum moving in (and interacting with) the electromagnetic field $F_{ij}$, it is then convenient to describe the latter by its projections on that local 3-plane \cite{Lichnerowicz}:
\begin{equation}
{\mathcal E}_i := F_{ik}u^k,\qquad {\mathcal B}^i := {\frac {1}{2}}\epsilon^{ijk}F_{jk}.\label{EBvec4}
\end{equation}
The field strength is uniquely reconstructed from these 4-vectors as
\begin{equation}\label{FEB}
F_{ij} = {\frac 1 {c^2}}\left({\mathcal E}_iu_j - {\mathcal E}_ju_i\right) - \epsilon_{ijk}{\mathcal B}^k.
\end{equation}
In the rest frame comoving with the fluid, $u^i = \delta^i_0$, the 4-vectors ${\mathcal E}_i$ and ${\mathcal B}^i$ reduce to the three-dimensional electric and magnetic fields, $\bm{E}$ and $\bm{B}$, respectively. Analogously, we can construct the decomposition of the excitation tensor
\begin{equation}
H^{ij} = {\mathcal D}^ju^i - {\mathcal D}^iu^j + \epsilon^{ijk}{\mathcal H}_k\label{HDH}
\end{equation}
in terms of its projections
\begin{equation}\label{DHvec4}
{\mathcal D}^i := {\frac 1{c^2}}H^{ki}u_k,\qquad {\mathcal H}_i := -\,{\frac {1}{2}}\epsilon_{ijk}H^{jk}.
\end{equation}
They reduce in the rest-frame to the three-dimensional electric and magnetic excitations, $\bm{D}$ and $\bm{H}$, respectively.
  
Substituting (\ref{DHvec4}) and (\ref{EBvec4}) into the Maxwell equations (\ref{Max4}), and making use of $u^i$ and $P^i_j$ to project the result into ``time'' and ``space'', we find the electrodynamics of moving media in a ``three-dimensional'' disguise:
\begin{align}
\partial_i{\mathcal D}^i &= J_{\|} - {\frac 1{c^2}}a_i{\mathcal D}^i + {\frac 1{c^2}}({\rm curl}\,u)^i
{\mathcal H}_i,\label{maxP1}\\
-\,\dot{\mathcal D}^i + ({\rm curl}\,{\mathcal H})^i &= J_{\perp}^i + {\frac 1{c^2}}({\mathcal H}\times a)^i  
- {\frac 12}({\mathcal D}\times {\rm curl}\,u)^i - h^i{}_j{\mathcal D}^j + {\frac 23}{\mathcal D}^i\partial_ku^k,\label{maxP2}\\
\partial_i{\mathcal B}^i &= -\,{\frac 1{c^2}}a_i{\mathcal B}^i + {\frac 1{c^2}}({\rm curl}\,u)^i
{\mathcal E}_i,\label{maxP3}\\
\dot{\mathcal B}^i + ({\rm curl}\,{\mathcal E})^i &= {\frac 1{c^2}}({\mathcal E}\times a)^i + {\frac 12}
({\mathcal B}\times {\rm curl}\,u)^i + h^i{}_j{\mathcal B}^j - {\frac 23}{\mathcal B}^i\partial_ku^k.\label{maxP4}
\end{align}
Here $J_{\|} = {\frac 1{c^2}}u_iJ^i$ and $J_{\perp}^i = P^i_jJ^j$ are the ``time'' (= longitudinal) and ``space'' (= transversal) projections of the electric current. The additional terms on the right-hand sides depend on the acceleration $a^i = \dot{u}^i$, vorticity $({\rm curl}\,u)^i$, as well as the shear $h_{ij} = P^k_iP^l_j\partial_{(k}u_{l)} - {\frac 13}P_{ij}\partial_ku^k$ and the volume expansion $\partial_ku^k$ of the 4-velocity congruence, manifesting the non-inertial character of observer's reference frame comoving with matter in an arbitrary way. 

To make the electromagnetic theory completed, we need the constitutive relation. Decomposing the constitutive tensor $\chi^{ijkl}$ into ``space'' and ``time'' parts, we find the three projections
\begin{equation}
\varepsilon^{ij} := -\,\mu_0{\frac {1}{c^2}}u_ku_l\chi^{ikjl},\qquad
\mu^{-1}_{ij} := \mu_0{\frac 14}\epsilon_{ikl}\epsilon_{jmn}\chi^{klmn},\qquad
\gamma_i{}^j := {\frac 1{2c^2}}\epsilon_{imn}u_k\chi^{jkmn}.\label{mega}
\end{equation}
The following properties are obvious: $\varepsilon^{ij} = \varepsilon^{ji}$, $\mu^{-1}_{ij} = \mu^{-1}_{ji}$,  $\varepsilon^{ij}u_j = 0$, $\mu^{-1}_{ij}u^j = 0$, $\gamma_i{}^ju_j = 0$, and $\gamma_i{}^ju^i = 0$. One can check that the constitutive tensor is uniquely reconstructed from these three pieces \cite{Lichnerowicz,Balakin:2005}:
\begin{align}
\chi^{ijkl} =&\, -\,\varepsilon_0\,(u^iu^k\varepsilon^{jl} - u^ju^k\varepsilon^{il} - u^iu^l\varepsilon^{jk}
+ u^ju^l\varepsilon^{ik}) + \mu^{-1}_0\mu^{-1}_{mn}\epsilon^{ijm}\epsilon^{kln}\nonumber\\
&\,+\, \epsilon^{ijm}(u^k\gamma_m{}^l - u^l\gamma_m{}^k)
+ \epsilon^{klm}(u^i\gamma_m{}^j - u^j\gamma_m{}^i).\label{chiP}
\end{align}
It is worthwhile to notice that the constitutive tensor has exactly the same algebraic properties as the Riemann curvature tensor in general relativity theory. A similar to (\ref{chiP}) representation of the curvature is known as the Bel decomposition \cite{Bel:1958,Costa:2014}, and it plays important role in the analysis of various gravitational effects \cite{Jordan:1960,Hawking:1966}.

Substituting (\ref{chiP}), (\ref{EBvec4}) and (\ref{DHvec4}) into (\ref{CR}), we recast the constitutive relation into a familiar ``three-dimensional'' disguise:
\begin{align}
{\mathcal D}^i &= \varepsilon_0\varepsilon^{ij}{\mathcal E}_j + \gamma_j{}^i{\mathcal B}^j,\label{CR1a}\\
{\mathcal H}_i &= \mu^{-1}_0\mu^{-1}_{ij}{\mathcal B}^j - \gamma_i{}^j{\mathcal E}_j.\label{CR2a}
\end{align}
When the material medium is at rest, one recovers (\ref{CR1})-(\ref{CR2}).

As a result, for the electromagnetic Lagrangian (\ref{Le}) we derive
\begin{equation}
L^{\rm e} = {\frac 12}\left({\mathcal D}^i{\mathcal E}_i - {\mathcal B}^i{\mathcal H}_i\right) = 
{\frac 12}\left(\varepsilon_0\varepsilon^{ij}{\mathcal E}_i{\mathcal E}_j - \mu^{-1}_0\mu^{-1}_{ij}{\mathcal B}^i
{\mathcal B}^j\right) + \gamma_i{}^j{\mathcal B}^i{\mathcal E}_j.\label{Lee}
\end{equation}

We are now in a position to clarify the structure of the energy-momentum tensors. Substituting (\ref{EBvec4}) and (\ref{DHvec4}) into (\ref{TM}), we find for the Minkowski energy-momentum tensor:
\begin{equation}
{\stackrel {\rm M}\Sigma}{}_k{}^i = {\mathcal E}_k{\mathcal D}^i + {\mathcal H}_k{\mathcal B}^i 
+ \Bigl({\frac 12}\delta_k^i - P_k^i\Bigr)\,({\mathcal D}^j{\mathcal E}_j + {\mathcal B}^j{\mathcal H}_j)
+ u^i\left({\mathcal D}\times{\mathcal B}\right)_k + {\frac 1{c^2}}u_k\left({\mathcal E}\times{\mathcal H}
\right)^i,\label{TM1}
\end{equation}
whereas using (\ref{EBvec4}) and (\ref{DHvec4}) in (\ref{TA}), we derive the Abraham energy-momentum tensor:
\begin{equation}\label{TA1}
{\stackrel {\rm A}\Sigma}{}_k{}^i = {\frac 12}({\mathcal E}_k{\mathcal D}^i + {\mathcal H}_k{\mathcal B}^i + 
{\mathcal E}^i{\mathcal D}_k + {\mathcal H}^i{\mathcal B}_k)  
+ \Bigl({\frac 12}\delta_k^i - P_k^i\Bigr)\,({\mathcal D}^j{\mathcal E}_j + {\mathcal B}^j{\mathcal H}_j)
+ {\frac 1{c^2}}u^i({\mathcal E}\times{\mathcal H})_k + {\frac 1{c^2}}u_k({\mathcal E}\times{\mathcal H})^i.
\end{equation}
Therefrom, we have the {\it longitudinal projections}
\begin{align}
{\frac 1{c^2}}u_iu^k {\stackrel {\rm M}\Sigma}{}_k{}^i &= {\frac 1{c^2}}u_iu^k {\stackrel {\rm A}\Sigma}{}_k{}^i
= {\frac 12}\,({\mathcal D}^j{\mathcal E}_j + {\mathcal B}^j{\mathcal H}_j),\\
{\frac 1{c^2}}u^k {\stackrel {\rm M}\Sigma}{}_k{}^i &= {\frac 1{c^2}}u^k {\stackrel {\rm A}\Sigma}{}_k{}^i 
= {\frac 1{2c^2}}\,({\mathcal D}^j{\mathcal E}_j + {\mathcal B}^j{\mathcal H}_j)\,u^i
+ {\frac 1{c^2}}({\mathcal E}\times{\mathcal H})^i,\\
{\frac 1{c^2}}u_i {\stackrel {\rm M}\Sigma}{}_k{}^i &= {\frac 1{2c^2}}\,({\mathcal D}^j{\mathcal E}_j
+ {\mathcal B}^j{\mathcal H}_j)\,u_k + ({\mathcal B}\times{\mathcal D})_k,\\
{\frac 1{c^2}}u_i {\stackrel {\rm A}\Sigma}{}_k{}^i &= {\frac 1{2c^2}}\,({\mathcal D}^j{\mathcal E}_j
+ {\mathcal B}^j{\mathcal H}_j)\,u_k + {\frac 1{c^2}}({\mathcal E}\times{\mathcal H})_k.
\end{align}
This is completely consistent with (\ref{UM})-(\ref{pM}) and (\ref{UA})-(\ref{sA}). Finally, the components of the stress tensors arise as {\it transversal projections}, cf. (\ref{MM}) and (\ref{AA}):
\begin{align}
{\stackrel {\rm M}\Sigma}_\perp{}_k{}^i = P_k^mP^i_n\,{\stackrel {\rm M}\Sigma}{}_m{}^n &=
{\mathcal E}_k{\mathcal D}^i + {\mathcal H}_k{\mathcal B}^i 
- {\frac 12}\,P_k^i\,({\mathcal D}^j{\mathcal E}_j + {\mathcal B}^j{\mathcal H}_j),\\
{\stackrel {\rm A}\Sigma}_\perp{}_k{}^i = P_k^mP^i_n\,{\stackrel {\rm A}\Sigma}{}_m{}^n &=
{\frac 12}({\mathcal E}_k{\mathcal D}^i + {\mathcal H}_k{\mathcal B}^i + {\mathcal E}^i{\mathcal D}_k +
{\mathcal H}^i{\mathcal B}_k) - {\frac 12}\,P_k^i\,({\mathcal D}^j{\mathcal E}_j + {\mathcal B}^j{\mathcal H}_j).
\end{align}

\section{Relativistic spinning fluid model}\label{weyss}

In order to analyse the Abraham-Minkowski issue for the case of complex moving media, we need an appropriate description of matter, and the field-theoretic variational formalism appears to provide the most convenient framework. A simple medium is usually modeled as an ideal fluid, the elements of which are structureless particles (i.e., no spin or other internal degrees of freedom are present). Such a continuous medium (e.g., see \cite{taub,schutz,schutz2,bailyn} for the relevant earlier work, as for the general discussion of the relativistic ideal fluids readers should refer to \cite{anile1,Rezzolla}) is characterized in the Eulerian approach by the fluid 4-velocity $u^i$, the {\it internal energy} density $\rho = \rho(\nu, s)$, the {\it particle density} $\nu$, the {\it entropy} $s$, and the {\it identity (Lin) coordinate} $X$ \cite{lin}. Furthermore, we assume that the motion of a fluid is such that the number of particles is constant and that the entropy and the identity of the elements is conserved. Due to the conservation of the entropy only reversible processes are allowed. In mathematical terms this means that the following constraints are imposed on the variables:
\begin{equation}
\partial_i(\nu u^i) = 0,\qquad u^i\partial_i X = 0,\qquad u^i\partial_i s = 0.\label{l3}
\end{equation}

Continuous media with {\it microstructure} \cite{capriz,corben,coss,halb,maugin,Eringen99,TT} are characterized by additional variables describing internal properties of fluid's elements. As a first fundamental examples of matter with microstructure, we now consider the spinning fluid \cite{Kopc,sf1,sf2,ray72,WR}.

\subsection{Lagrangian of the spinning fluid}

Following the Cosserat approach \cite{coss}, matter with microstructure is described as a continuous medium the elements of which are characterized by the 4-velocity $u^i$ and the material triad $b^i_A$, $A=1,2,3$. The latter is assumed to be rigid, which means that angles between triad's legs and the velocity are constant and subject to the orthogonality and normalization conditions:
\begin{align}\label{ort}
g_{ij}u^iu^j = c^2,\qquad g_{ij}u^ib^j_A = 0,\qquad g_{ij}b^i_Ab^j_B = g_{AB} = -\delta_{AB}.
\end{align}
Taken together, these variables comprise the {\it material frame} attached to elements of the fluid,
\begin{equation}
e^i_\alpha = \left\{u^i, b^i_{\hat{1}}, b^i_{\hat{2}}, b^i_{\hat{3}}\right\}.\label{matframe}
\end{equation}
The inverse coframe is constructed as
\begin{equation}
e_i^\alpha = \Bigl\{{\frac 1{c^2}}u_i, b_i^{\hat{1}}, b_i^{\hat{2}}, b_i^{\hat{3}}\Bigr\}\,,\label{matcof}
\end{equation}
where we introduced the co-triad by $b_i^A := g_{ij}g^{AB}b^j_B$. By definition, we have from (\ref{ort})
\begin{align}
b^i_A b_i^B = \delta_A^B,\qquad b^i_A b_j^A = P^i_j = \delta_j^i - {\frac 1{c^2}}\,u^iu_j,\label{dij}
\end{align}
and one can verify that $e_i^\alpha e^j_\alpha = \delta_i^j$. Here we recover the projector (\ref{pro}) on the local three-dimensional hyperplane, spanned by the triad $b^i_A$,  orthogonal to the 4-velocity. 

The evolution of the material frame is encoded in the generalized acceleration tensor:
\begin{equation}
\Omega^\alpha{}_\beta := e^\alpha_i u^k\partial_k e^i_\beta.\label{Oab}
\end{equation}
Its components encompass fluid's acceleration $\Omega^{\hat{0}}{}_A = -\,b^i_A\,u^k\partial_ku_i$ and an angular velocity 
\begin{align}
\Omega_{AB} = g_{ij}b^i_Au^k\partial_kb^j_B,\label{Om}
\end{align}
measured by an observer comoving with the fluid. 

After these preliminaries, we can formulate the spinning fluid model as follows. The physical properties of matter are characterized by the internal energy density $\rho(\nu, s, \mu^{AB})$, particle number density $\nu$, specific spin density $\mu^{AB} = -\mu^{BA}$, the entropy $s$, and the identity Lin coordinate $X$. As usual, we assume the standard thermodynamics encoded in the Gibbs law
\begin{align}
Tds = d\left({\frac {\rho}{\nu}}\right) + p\,d\left({\frac {1}{\nu}}\right)
- {\frac 12}\omega_{AB}\,d\mu^{AB},\label{gibbs}
\end{align}
with $T$ temperature, $p$ pressure, and $\omega_{AB}$ conjugate to $\mu^{AB}$. 

The Lagrangian then reads
\begin{equation}\label{L}
L^{\rm m} = -\,\rho - {\frac 12}\nu\mu^{AB}\Omega_{AB} - {\frac 12}\,\zeta\nu\mu^{AB}b^i_Ab^j_BF_{ij} + L^{\rm c}.
\end{equation}
Here the first two terms describe the potential and kinetic (spin-rotation) energy contributions, whereas the third term is responsible for the Pauli type interaction of the magnetic dipole moment density with the electromagnetic field. The corresponding coupling constant has the dimension $[\zeta] = {\frac {[{\rm electric\ charge}]}{[{\rm mass}]}}$, so that $[\zeta\mu^{AB}]$ has the dimension of the magnetic dipole moment, since $[\mu^{AB}] = [\hbar]$. Finally, the last term collects all the constraints imposed on fluid's variables by means of the Lagrange multipliers:
\begin{equation}
L^{\rm c} = -\,\nu u^i\partial_i\lambda_1 + \lambda_2 u^i\partial_iX + \lambda_3 u^i\partial_is 
+ \lambda_0(g_{ij}u^iu^j - c^2) + \lambda^A g_{ij}u^ib^j_A + \lambda^{AB}(g_{ij}b^i_Ab^j_B - g_{AB}).\label{Lc}
\end{equation}

\subsection{Euler-Lagrange equations}

Variation with respect to the Lagrange multipliers $\lambda^{AB}, \lambda^A, \lambda_0, \lambda_1, \lambda_2, \lambda_3$ yields (\ref{l3}) and (\ref{ort}), in other words, we recover the orthogonality and normalization constrains, together with the conservation of the entropy, number and identity of particles during fluid's motion.

Variations with respect to the fluid variables $X, s, \nu$ yield, respectively:
\begin{align}
{\frac {\delta L^{\rm m}}{\delta X}} &= -\,\partial_i(\lambda_2 u^i) = 0,\label{vX}\\
{\frac {\delta L^{\rm m}}{\delta s}} &= -\,\partial_i(\lambda_3 u^i) - \nu T = 0,\label{vs}\\
{\frac {\delta L^{\rm m}}{\delta \nu}} &= -\,u^i\partial_i\lambda_1 - {\frac {\rho + p}{\nu}} 
- {\frac 12}\mu^{AB}\Omega_{AB} - {\frac 12}\,\zeta\mu^{AB}b^i_Ab^j_BF_{ij}.\label{vrho}
\end{align}
In addition, varying the action with respect to the fluid velocity $u^i$, we find
\begin{align}
{\frac {\delta L^{\rm m}}{\delta u^i}} &= \lambda^Ag_{ij}b^j_A + 2\lambda_0u_i - \nu\partial_i\lambda_1
+ \lambda_2\partial_iX + \lambda_3\partial_is - {\frac 12}\nu\mu^{AB}g_{kl}b^k_A\partial_ib^l_B .\label{vu} 
\end{align}
Finally, the variation with respect to the fluid triad $b^i_A$ yields
\begin{align}
2\lambda^{AB}g_{ij}b^j_B + \lambda^Au_i - \nu\mu^{AB}g_{ij}u^k\partial_kb^j_B 
- \,{\frac 12}\nu b^j_Bg_{ij}u^k\partial_k\mu^{AB} - \zeta\nu\mu^{AB}b^j_BF_{ij} = 0,\label{vb}
\end{align}
whereas the variation with respect to the fluid spin $\mu^{AB}$ results in 
\begin{align}
\omega_{AB} + \Omega_{AB} + \zeta b^i_Ab^j_BF_{ij} = 0.\label{vmu}
\end{align}

Contracting (\ref{vu}) with $u^i$ and using (\ref{vrho}), we derive
\begin{align}
2\lambda_0c^2 = -\,\rho - p - {\frac 12}\,\zeta\nu\mu^{AB}b^i_Ab^j_BF_{ij}
- \nu{\frac {\delta L^{\rm m}}{\delta \nu}} + u^i{\frac {\delta L^{\rm m}}{\delta u^i}}.\label{l0s} 
\end{align}
Similarly, contracting (\ref{vb}) with $u^i$: 
\begin{align}
\lambda^A = {\frac {\nu}{c^2}}\mu^{AB}u_iu^k\partial_kb^i_B + \zeta\nu\mu^{AB}u^ib^j_BF_{ij}.\label{lAs}
\end{align}
Next, contracting (\ref{vb}) with $b^i_C$, we obtain
\begin{align}\label{lAB1}
2\lambda^{AB} - \nu\mu^{AC}\Omega^B{}_C - {\frac 12}\nu u^k\partial_k\mu^{AB} - \zeta\nu\mu^{AC}F^B{}_C= 0.
\end{align}
Here $F_{AB} = b^i_Ab^j_BF_{ij}$ and the indices $A,B,\dots$ are raised and lowered with the help of $g_{AB} = -\delta_{AB}$, in particular, $F^A{}_C = g^{AB}F_{BC}$ The symmetric part determines the Lagrange multiplier
\begin{align}
\lambda^{AB} = {\frac 12}\nu\mu^{(A|C|}\Omega^{B)}{}_C + {\frac 12}\zeta\nu\mu^{(A|C|}F^{B)}{}_C .\label{lAB2}
\end{align}
The antisymmetric part yields the equation of motion of spin
\begin{align}\label{dmu}
u^k\partial_k\mu^{AB} + \mu^{CB}\Omega^{A}{}_C + \mu^{AC}\Omega^{B}{}_C
+ \zeta\mu^{CB}F^{A}{}_C + \zeta\mu^{AC}F^{B}{}_C = 0.
\end{align}

\subsection{Canonical Noether currents of spin and energy-momentum}

We now define the tensor of spin density
\begin{align}
{\mathcal S}^{ij} = {\frac 12}\nu b^i_Ab^j_B\mu^{AB}.\label{spin}
\end{align}
Contracting (\ref{dmu}) with ${\frac 12}\nu b^i_Ab^j_B$, we find the covariant 
equation of motion of spin
\begin{align}\label{dSij}
\dot{\mathcal S}^{ij} - {\frac {u^iu_k}{c^2}}\dot{\mathcal S}^{kj} - {\frac {u^ju_k}{c^2}}\dot{\mathcal S}^{ik}
+ \zeta\left({\mathcal S}^{ik}P^j_l - {\mathcal S}^{jk}P^i_l\right)F^l{}_k = 0. 
\end{align}
Here $\dot{\mathcal S}^{ij} = \partial_k\left(u^k{\mathcal S}^{ij}\right)$. 

As a result, making use of the standard Euler-Lagrange machinery \cite{hehl76,OR1,OR2} for the spinning fluid Lagrangian (\ref{L}), we derive the canonical energy-momentum and spin tensors: 
\begin{align}
\sigma_k{}^i =& \,u^i {\mathcal P}_k - p^{\rm eff}\left(\delta_k^i - {\frac {u^iu_k}{c^2}}\right)
- 2\zeta F_{kj}{\mathcal S}^{ij},\label{Sig}\\
S_{ij}{}^k =& \,u^k {\mathcal S}_{ij}.\label{tau}
\end{align}
Here we denoted
\begin{align}\label{Pk}
{\mathcal P}_k = {\frac 1{c^2}}\left[u_k\left(\rho + \zeta{\mathcal S}^{ij}F_{ij}\right) - 2u^l
\dot{\mathcal S}_{kl} - 2\zeta u^iF_{ij}{\mathcal S}_k{}^j\right] + P^i_k{\frac {\delta L^{\rm m}}{\delta u^i}},
\end{align}
where the effective pressure picks up an additional term
\begin{align}
p^{\rm eff} = p + \nu{\frac {\delta L^{\rm m}}{\delta \nu}}.\label{peff}
\end{align}
It is important to verify the angular momentum balance law:
\begin{align}
\sigma_{[ij]} + \partial_k S_{ij}{}^k = {\frac {\delta L^{\rm m}}{\delta u^{[i}}}u_{j]}.\label{tskew}
\end{align}
Indeed, by making use (\ref{Sig})-(\ref{Pk}) we can explicitly demonstrate that (\ref{tskew}) is a consequence of the equations of motion of spin (\ref{dSij}). 

In a similar way, we can verify that the canonical energy-momentum satisfies the balance law
\begin{align}\label{q2b}
\partial_i\sigma_k{}^i = -\,{\frac {\delta L^{\rm m}}{\delta\nu}}\,\partial_k\nu - {\frac {\delta L^{\rm m}}
{\delta u^i}}\,\partial_ku^i,
\end{align}
in view of the Euler-Lagrange equations.

\subsection{Electromagnetic Lagrangian}

The structure of the energy and momentum of matter, as well as the corresponding balance laws of the angular momentum and the energy-momentum currents, are still incomplete since the equations above contain variational derivatives with respect to fluid's particle density $\nu$ and velocity $u^i$. In order to fix this deficiency, we have to recall that the spinning fluid is an {\it open system} that interacts with the electromagnetic field. In technical terms, this means that we need to consider the {\it total action} $I = {\frac 1c}\int d^4x\left(L^{\rm m} + L^{\rm e}\right)$ for the sum of the matter Lagrangian (\ref{L}) and the electromagnetic Lagrangian (\ref{Le}), and to use the field equations for the {\it closed system},
\begin{equation}\label{eqnu}
{\frac {\delta L^{\rm m}}{\delta\nu}} + {\frac {\delta L^{\rm e}}{\delta\nu}} = 0,\qquad
{\frac {\delta L^{\rm m}}{\delta u^i}} + {\frac {\delta L^{\rm e}}{\delta u^i}} = 0.
\end{equation}
In order to demonstrate how this works, we now specialize to the case of a moving isotropic medium which is described by the Lagrangian
\begin{equation}
L^{\rm e} = -\,{\frac 1{4\mu_0\mu}}\,g_{\rm opt}^{ij}\,g_{\rm opt}^{kl}\,F_{ik}F_{jl} =
-\,{\frac 1{4\mu_0\mu}}\left[F_{ij}F^{ij} + 2{\frac {n^2 -1}{c^2}}\,F_{ik}u^kF^{il}u_l\right],\label{Lem}
\end{equation}
where the so-called {\it optical metric} was first introduced by Gordon \cite{wgordon}:
\begin{equation}
g_{\rm opt}^{ij} = g^{ij} - {\frac {1 - n^2}{c^2}}\,u^i\,u^j.\label{gopt}
\end{equation}
Here $n = \sqrt{\varepsilon\mu}$ is the refractive index of the medium. The corresponding constitutive tensor thus reads
\begin{equation}\label{chi1}
\chi{}^{ijkl} = {\frac 1{\mu_0\mu}}\left(g_{\rm opt}^{ik}g_{\rm opt}^{jl} - g_{\rm opt}^{il}g_{\rm opt}^{jk}\right).
\end{equation}
Computing the electromagnetic excitation (\ref{Hij}) for the Lagrangian (\ref{Lem}) we find
\begin{equation}
H^{ij} = {\frac 1{\mu_0\mu}}\,g_{\rm opt}^{ik}\,g_{\rm opt}^{jl}\,F_{kl} = {\frac 1{\mu_0\mu}}
\left[F^{ij} + {\frac {n^2 - 1}{c^2}}\left(F^{ik}u_ku^j - F^{jk}u_ku^i\right)\right].\label{Hij0}
\end{equation}

The Lagrangian (\ref{Lem}) depends on fluid's velocity $u^i$ via the Gordon metric (\ref{gopt}), and it also depends on $\nu$ when we take into account the possible {\it electrostriction and magnetostriction} effects, and allow for the permittivity $\varepsilon$ and the permeability $\mu$ to be the functions of the particle density, 
\begin{equation}
\varepsilon = \varepsilon(\nu),\qquad \mu = \mu(\nu).\label{epsmu}
\end{equation}

In terms of the electric and magnetic 4-vectors (\ref{EBvec4}), the Lagrangian (\ref{Lem}) can be recast into a nice and compact form
\begin{equation}\label{Lem2}
L^{\rm e} = -\,{\frac 12}\left(\varepsilon\varepsilon_0{\mathcal E}^2 - {\frac {{\mathcal B}^2}{\mu\mu_0}}\right),
\end{equation}
with the obvious abbreviations ${\mathcal E}^2 = {\mathcal E}_i{\mathcal E}^i$ and ${\mathcal B}^2 = {\mathcal B}_i{\mathcal B}^i$. Since both vectors are by construction orthogonal to the 4-velocity, ${\mathcal E}_iu^i = 0$ and ${\mathcal B}^iu_i = 0$, we have ${\mathcal E}^2 \leq 0$ and ${\mathcal B}^2 \leq 0$. Moreover, the constitutive relation (\ref{Hij0}) assumes a clear structure in terms of the excitation 4-vectors (\ref{DHvec4}):
\begin{equation}
{\mathcal D}^i = -\,\varepsilon\varepsilon_0g^{ij}{\mathcal E}_j,\qquad
{\mathcal H}_i = -\,{\frac 1{\mu\mu_0}}g_{ij}{\mathcal B}^j.\label{CRopt}
\end{equation}
For the medium at rest, this reduces to (\ref{CRsimple}).

The variational derivatives of the electromagnetic Lagrangian with respect to the material variables are easily computed:
\begin{eqnarray}
{\frac {\delta L^{\rm e}}{\delta u^i}} &=& -\,{\frac {n^2 -1}{\mu\mu_0c^2}}F_{ki}F^{kl}u_l,\label{dLeu}\\
{\frac {\delta L^{\rm e}}{\delta \nu}} &=& -\,{\frac 12}\left(\varepsilon_0
{\frac {\partial\varepsilon}{\partial\nu}}\,{\cal E}^2 + {\frac 1{\mu_0
\mu^2}}{\frac {\partial\mu}{\partial\nu}}\,{\cal B}^2\right).\label{dLen}
\end{eqnarray}
Using the equations of motion (\ref{eqnu}), we can find the explicit form of the canonical energy-momentum of medium, too. In particular, combining (\ref{eqnu}) with (\ref{dLeu}) and (\ref{dLen}), we have
\begin{eqnarray}
{\cal P}_k &=& {\frac 1{c^2}}\left[u_k(\rho + \zeta{\mathcal S}^{ij}F_{ij}) - 2u^l\dot{\mathcal S}_{kl}
- 2\zeta u^iF_{ij}{\mathcal S}_k{}^j + {\frac {n^2 - 1}{\mu\mu_0}}\left(
F_{ik}F^{il}u_l - {\frac {u_k}{c^2}}\,F_{jn}u^nF^{jl}u_l\right)\right],\label{PkF}\\
p^{\rm eff} &=& p + {\frac 12}\nu\left(\varepsilon_0{\frac {\partial\varepsilon}
{\partial\nu}}\,{\cal E}^2 + {\frac 1{\mu_0\mu^2}}{\frac {\partial\mu}
{\partial\nu}}\,{\cal B}^2\right).\label{presseff}
\end{eqnarray}
Quite remarkably, the effective pressure describes the electro- and magnetostriction effects. Substituting (\ref{PkF}) into (\ref{Sig}), we obtain the final expression for the energy-momentum tensor of matter
\begin{eqnarray}\label{emtF0}
\sigma_k{}^i = {\frac \rho {c^2}}u_ku^i - p^{\rm eff}\left(\delta_k^i - {\frac {u_ku^i}{c^2}}\right)
+ {\stackrel {\rm s}\sigma}{}_k{}^i + {\frac {n^2 - 1}{\mu\mu_0c^2}}\left( u^iF_{nk}F^{nl}u_l
- {\frac {u_ku^i}{c^2}}\,F_{jn}u^nF^{jl}u_l\right),
\end{eqnarray}
where all the spin contributions are collected in
\begin{equation}\label{SigS}
{\stackrel {\rm s}\sigma}{}_k{}^i = {\frac {u^i}{c^2}}\left[\zeta u_k{\mathcal S}^{ij}F_{ij}
- 2u^l\dot{\mathcal S}_{kl} - 2\zeta u^iF_{ij}{\mathcal S}_k{}^j\right] - 2\zeta F_{kj}{\mathcal S}^{ij}.
\end{equation}
It is worthwhile to notice that this energy-momentum tensor satisfies the angular momentum conservation law, as if the fluid is a closed system
\begin{align}
{\stackrel {\rm s}\sigma}{}_{[ij]} + \partial_k S_{ij}{}^k = 0.\label{Sc}
\end{align}

\subsection{Minkowski and Abraham energy-momentum tensors}

Substituting the excitation (\ref{Hij0}) into (\ref{TM}), we find the explicit form of the canonical (=Minkowski) energy-momentum of the electromagnetic field:
\begin{equation}
{\stackrel {\rm M} \Sigma}{}_k{}^i = {\frac 1{\mu\mu_0}}\bigg[-\,F^{ij}F_{kj} + {\frac 1{4}}
F^{jl}F_{jl}\,\delta_k^i + {\frac {n^2 - 1}{c^2}}\left(F_{kn}F^{nl}u_lu^i - F_{kl}u^l
F^{in}u_n + {\frac {1}{2}}F_{nl}u^lF^{nj}u_j\delta_k^i\right)\bigg].\label{emtEF}
\end{equation}
This tensor is not symmetric, and its antisymmetric part reads explicitly
\begin{equation}\label{emtMskew}
{\stackrel {\rm M} \Sigma}{}_{[jk]} = -\,{\frac {n^2 - 1}{\mu\mu_0c^2}}\,u_{[j}F_{k]i}F^{il}u_l.
\end{equation}
On the other hand, substituting (\ref{Hij0}) into (\ref{TA}), we obtain the Abraham energy-momentum tensor
\begin{equation}
{\stackrel {\rm A} \Sigma}{}_k{}^i = {\frac 1{\mu\mu_0}}\bigg[-\,F^{ij}F_{kj} + {\frac 1{4}}F^{jl}F_{jl}\,\delta_k^i
+ {\frac {n^2 - 1}{c^2}}\left(- F_{kl}u^lF^{in}u_n - {\frac {u_ku^i}{c^2}}\,F_{jn}u^nF^{jl}u_l
+ {\frac {1}{2}}F_{nl}u^lF^{nj}u_j\delta_k^i\right)\bigg].\label{emtA0}
\end{equation}

The total canonical energy-momentum tensor of the coupled system (spinning fluid + electromagnetic field) is the sum of the electromagnetic (Minkowski) tensor (\ref{emtEF}) and of the energy-momentum of the fluid (\ref{emtF0}). After some simple algebra, by comparing the formula (\ref{emtA0}) with (\ref{emtEF}), we establish a remarkable relation between the energy-momentum tensors:
\begin{equation}\label{Ttot}
\sigma_k{}^i + {\stackrel {\rm M} \Sigma}{}_k{}^i = {\frac \rho {c^2}}u_ku^i - p^{\rm eff}\left(\delta_k^i
- {\frac {u^iu_k}{c^2}}\right) + {\stackrel {\rm s}\sigma}{}_k{}^i + {\stackrel {\rm A} \Sigma}{}_k{}^i.
\end{equation}
As we see, the Abraham tensor arises as a peculiar patchwork from the Minkowski canonical tensor of the electromagnetic field and from the canonical energy-momentum of the matter by ``absorbing" all the terms which explicitly contain the electromagnetic field, except for the electrostriction and magnetostriction terms. It is reasonable to identify the first three terms on the right-hand side of (\ref{Ttot}) as a {\it kinetic energy-momentum tensor} of matter:
\begin{equation}\label{kinS}
\kappa_k{}^i := {\frac \rho {c^2}}u_ku^i - p^{\rm eff}\left(\delta_k^i
- {\frac {u^iu_k}{c^2}}\right) + {\stackrel {\rm s}\sigma}{}_k{}^i.
\end{equation}
It is not symmetric, due to a nontrivial spin, with the angular momentum conserved in a usual way
\begin{align}
\kappa_{[ij]} + \partial_k S_{ij}{}^k = 0.\label{kinA}
\end{align}

\section{Relativistic liquid crystal model}\label{crystal}

Let us now turn to the discussion of a second physically important example of a complex matter with microstructure: the liquid crystal medium \cite{degennes,dejeu,chandra,blinov}.

In contrast to the spinning fluid, miscrostructural properties of which are encoded in the Cosserat triad $b^i_A$, the microstructure of a liquid crystal, as a medium with uniaxial anisotropic properties, is described by the {\it director} 4-vector $N^i$ which is orthogonal to the 4-velocity
\begin{align}
N^{i}u_{i} = 0.\label{connn1}
\end{align}
In other words, $N^i$ is a spacelike 4-vector. In addition, it is normalized
\begin{align}
N^iN_i = -\,1.\label{connn3}
\end{align}
Geometrically, this object can be viewed as an elastic 1-bein, which represents an important example of elastic vielbeins \cite{Volovik:2022}.

The non-relativistic liquid crystal theory is a well established subject; see, for instance, \cite{degennes,dejeu,stewart,erik5,erik6,leslie1,leslie2,erik2,erik3,erik4}. On the other hand, the Lagrangian approach for the study of the dynamics of this medium with microstructure was developed only recently in \cite{lisin1,lisin2} (although the variational methods were used in \cite{Virga} for the analysis of the equilibrium problems for the liquid crystals).

In order to construct the kinetic energy of liquid crystal, we define the angular 4-velocity of the director by
\begin{equation}
\omega^{i} := (N\times \dot{N})^i = \epsilon^{ijk}N_{j}\dot{N}_{k},\label{omega}
\end{equation}
where the convective ``time'' derivative is naturally
\begin{align}
\dot{N}^{i} = u^{j}\partial_{j}N^{i}.
\end{align}

A liquid crystal medium gives an example of a fluid with microstructure, represented by the director field attached to each element of the fluid. A relativistic variational model for this system can be constructed by extending the variational model of an ideal fluid described in \cite{obukhov1} with an account of the four-dimensional generalization of the liquid crystal elastic potential energy \cite{LQ}:
\begin{align}\label{VF}
{\mathcal V} = {\frac 12}K_1\left(\partial_{i}N^{i}\right)^2 + {\frac 12}K_2\left(
N^i({\rm curl}\,N)_i\right)^2 - {\frac 12}K_3\left(N\times({\rm curl}\,N)\right)^2.
\end{align}
Here the three constant parameters $K_1$, $K_2$ and $K_3$ are known as Frank's elastic constants (elastic moduli), which are all independent from each other and also positive. One usually calls $K_1$ splay, $K_2$ twist, and $K_3$ bend constants. The so-called saddle-splay boundary term can be omitted, since it is a total derivative that does not contribute to the equations of motion. For a typical nematic crystal, one has $K_1 = 2.3\times 10^{-12}\,$N, $K_2 = 1.5\times 10^{-12}\,$N, $K_3 = 4.8\times 10^{-12}\,$N, see \cite{degennes,Virga}. A different sign of the $K_3$ term in (\ref{Lfluid}) is explained by the fact that the 4-vector $\left(N\times({\rm curl}\,N)\right)_i = \epsilon_{ijk}N^{j}\epsilon^{kln}\partial_{l}N_{n}$ is spacelike, hence the square of its 4-length is negative.

Then, together with the internal energy and the kinetic energy of the medium, the Lagrangian reads \cite{LQ}
\begin{equation}\label{Lfluid}
L^{\rm m} = -\,\rho(\nu,s) - {\frac {{\mathcal J}\nu}2}\,\omega^{i}\omega_{i} - {\mathcal V} + L^{\rm c}.
\end{equation}
As before, $\nu$ is the particle number density of the liquid crystal viewed as the relativistic fluid, and $\rho(\nu,s)$ is the internal energy density as a function of $\nu$ and the entropy $s$, whereas ${\mathcal J}$ is the moment of inertia of a material element of this medium. The last term imposes a set of constraints by means of the Lagrange multipliers $\lambda_{I}$, $I=0, \dots,5$:
\begin{equation}\label{LQc}
L^{\rm c} = -\,\nu u^{i}\partial_{i}\lambda_1 + \lambda_2 u^{i}\partial_{i}X + \lambda_3 u^{i}\partial_{i}s 
+ \lambda_0(u^{i}u_{i}-c^2) + \lambda_4(N^{i}N_{i} + 1) + \lambda_5 u^{i}N_{i},
\end{equation}
cf. eq. (\ref{Lc}), which ensures the fulfillment of the conditions (\ref{connn1}), (\ref{connn3}) and the normalization $g_{ij}u^iu^j = c^2$ throughout all the dynamics of the nematic liquid crystal, in addition to the particle number continuity law, the conservation of entropy and identity of particles along each streamline of the fluid (\ref{l3}).

\subsection{Field equations}

Variation of the action with respect to the Lagrange multipliers and the variables $s, X$ yield the set of equations (\ref{connn1}), (\ref{connn3}), (\ref{l3}), and (\ref{vX}), (\ref{vs}). In addition, the variations with respect to the essential field variables $\nu$, $u^i$, and $N^i$ now give
\begin{align}
{\frac {\delta{L}^{\rm m}}{\delta\nu}} =& -\,{\frac {\mathcal J}2}\,\omega^2 
- {\frac {p + \rho}\nu} - u^i\partial_i\lambda_1,\label{dLm_dnu}\\
{\frac {\delta{L}^{\rm m}}{\delta u^i}} =& {\mathcal J}\nu\Bigl(-P^j{}_k\partial_i N^k
+ {\frac{1}{c^2}}\dot{N}_i u^j\Bigr)\dot{N}_j - {\frac {\partial{\cal V}}{\partial u^i}} 
+ 2\lambda_0 u_i - \nu\partial_i\lambda_1 + \lambda_2\partial_i X + \lambda_3\partial_i s
+ \lambda_5 N_i,\label{dLm_du}\\
{\frac {\delta{L}^{\rm m}}{\delta N^i}} =& \partial_k\left({\mathcal J}\nu\,u^k\,P^j{}_i\,\dot{N}_j\right)
- {\frac {\delta{\cal V}}{\delta N^i}} + 2\lambda_4N_i + \lambda_5u_i.\label{dLm_dN}
\end{align}
Importantly, these expressions are not zero since one still has to take into account the dependence of the electromagnetic Lagrangian $L^{\rm e}$ on these variables, and ultimately one should use the field equations for the closed ``matter+field'' system:
\begin{align}
{\frac {\delta{L}^{\rm m}}{\delta \nu}} + {\frac {\delta{L}^{\rm e}}{\delta \nu}} = 0,\qquad 
{\frac {\delta{L}^{\rm m}}{\delta u^i}} + {\frac {\delta{L}^{\rm e}}{\delta u^i}} = 0,\qquad 
{\frac {\delta{L}^{\rm m}}{\delta N^i}} + {\frac {\delta{L}^{\rm e}}{\delta N^i}} = 0.\label{EqNN}
\end{align}

Contracting (\ref{dLm_dN}) with $u^i$ and $N^i$, we find the Lagrange multipliers
\begin{equation}\label{Lambda5}
\lambda_5 = {\frac {u^i}{c^2}}\left[{\frac {\delta{L}^{\rm m}}{\delta N^i}}
- \partial_k({\mathcal J}\nu\,u^k\,P^j{}_i\,\dot{N}_j) + {\frac {\delta{\cal V}}{\delta N^i}}\right],\quad
2\lambda_4 = -\,N^i\left[{\frac {\delta{L}^{\rm m}}{\delta N^i}} - \partial_k({\mathcal J}\nu\,u^k\,P^j{}_i
\,\dot{N}_j) + {\frac {\delta{\cal V}}{\delta N^i}}\right],
\end{equation}
and substituting these back into (\ref{dLm_dN}), we obtain the field equation for the director field:
\begin{align}\label{EqN}
\pi_i{}^j\left[\partial_k\left({\mathcal J}\nu\,u^k\,P^l{}_j\,\dot{N}_l\right) - {\frac {\delta{\cal V}}
{\delta N^j}}\right] = \pi_i{}^j{\frac {\delta{L}^{\rm m}}{\delta N^j}}.
\end{align}
Here we introduced the projector on the 2-dimensional space orthogonal to both $N^{i}$ and $u^{i}$,
\begin{align}
\pi_i^j := \delta_i^j - {\frac{1}{c^2}}u_{i}u^{j} + N_{i}N^{j}.\label{pi}
\end{align}
It obviously has the properties $\pi^{i}{}_{j}\pi^{j}{}_{k}=\pi^{i}{}_{k}$, $\pi_i^ju^i = 0$, $\pi_i^jN^i = 0$, and $\det\pi^{ij} = 0$. Note that we cannot yet put equal zero the right-hand side of (\ref{EqN}); it still should be evaluated later from the variation of the electromagnetic part of the total Lagrangian. 

Contracting (\ref{dLm_du}) with $u^i$, we find another Lagrange multiplier:
\begin{align}
2\lambda_0c^2 = u^i{\frac {\delta{L}^{\rm m}}{\delta u^i}} - \nu{\frac {\delta{L}^{\rm m}}{\delta\nu}}
- \rho - p + {\mathcal J}\nu\Bigl(\frac{1}{2}\omega^2 - \Bigl(\frac{1}{c^2}u^i\dot{N}_i\Bigr)^2\Bigr)
+ u^i {\frac {\partial{\mathcal V}}{\partial u^i}}.\label{Lambda0}
\end{align}
Here again we cannot put equal zero the first two terms on the right-hand side of (\ref{Lambda0}); they also should be inserted from (\ref{EqNN}). Finally, let us notice that the material Lagrangian ``on-shell'' (i.e., after making use of the field equations) reads:
\begin{align}\label{onLm}
{L}^{\rm m} = p + \nu{\frac {\delta{L}^{\rm m}}{\delta\nu}} - {\mathcal V}.
\end{align}

\subsection{Canonical Noether currents for matter}

For the relativistic liquid crystal Lagrangian (\ref{Lfluid}), we derive the canonical energy-momentum tensors (by making use of the above findings):
\begin{align}
\sigma_k{}^i = {\frac {\rho}{c^2}}u_ku^i - p^{\rm eff}\left(\delta_k^i - {\frac {u^iu_k}{c^2}}\right)
+ {\stackrel{\rm s}\sigma}{}_k{}^i + u^i\biggl[P_k^j{\frac {\delta{L}^{\rm m}}{\delta u^j}}
- N_k{\frac {u^j}{c^2}}\,{\frac {\delta{L}^{\rm m}}{\delta N^j}}\biggr].\label{SigLQ}
\end{align} 
Here the microstructural part of the energy-momentum reads
\begin{equation}
{\stackrel{\rm s}\sigma}{}_k{}^i = u^i {\mathcal P}_k + {\stackrel{\rm F}\sigma}{}_k{}^i,\label{SigLQs}
\end{equation}
with the generalized 4-momentum density 
\begin{equation}
{\mathcal P}_k =  - \,{\frac {{\mathcal J}\nu\omega^2}{2c^2}}\,u_k - P_k^j\biggl[{\frac {{\mathcal J}\nu}{c^2}}
\dot{N}_j\,u^l\dot{N}_l - {\frac {\partial{\cal V}}{\partial u^j}}\biggr] + N_k{\frac {u^j}{c^2}}\biggl[
\partial_n({\mathcal J}\nu u^n P_j{}^l\dot{N}_l) - {\frac {\delta{\cal V}}{\delta N^j}}\biggr],\label{Pmom}
\end{equation}
and the last term in (\ref{SigLQs}) is the Frank elastic stress tensor
\begin{align}\label{Femt}
{\stackrel{\rm F}\sigma}{}_i{}^j :=&\, -\,{\frac {\partial{\cal V}}{\partial
\partial_j N^k}}\partial_i N^k + \delta_i^j\,{\cal V}\\
=&\,-\,K_1(\partial_k N^k)(\partial_i N^j) - K_2(\partial_iN_k)P^j_pP^k_q(\partial^pN^q-\partial^qN^p)
-(K_2 - K_3)(\partial_iN_k)N^jP_l^k(N^p\partial_pN^l)\nonumber\\
&\,+\,{\frac 12}K_1\delta_i^j(\partial_{k}N^k)^2 + {\frac 12}K_2\delta_i^j(\epsilon^{klm}N_{k}\partial_{l}N_{m})^2
- {\frac 12}K_3\delta_i^j(\epsilon_{pmk}N^{m}\epsilon^{kln}\partial_{l}N_{n})^2,\label{TFtot}
\end{align}
As before, the effective pressure picks up an additional term (\ref{peff}). 

The canonical spin density tensor of matter is also straightforwardly derived from (\ref{Lfluid}):
\begin{equation}
S_{ij}{}^k = - \,N_{[i}P^l_{j]}\Bigl[{\mathcal J}\nu u^k\dot{N}_l + K_2P^{kn}(\partial_n N_l - \partial_l N_n)
+ (K_2 - K_3)N^kN^n\partial_n N_l\Bigr] - K_1N_{[i}\,\delta_{j]}^k\partial_l N^l.\label{spin3}
\end{equation}

We can verify the balance laws of the canonical energy-momentum and the angular momentum:
\begin{align}
\partial_i\sigma_k{}^i &= -\,{\frac {\delta L^{\rm m}}{\delta\nu}}\,\partial_k\nu
-{\frac {\delta L^{\rm m}}{\delta u^i}}\,\partial_ku^i - {\frac {\delta L^{\rm m}}{\delta N^i}}\,\partial_kN^i,\\
\sigma_{[ij]} + \partial_k S_{ij}{}^k &= {\frac {\delta L^{\rm m}}{\delta u^{[i}}}u_{j]}
+ {\frac {\delta L^{\rm m}}{\delta N^{[i}}}N_{j]}.\label{tskewLQ}
\end{align}

\subsection{Electromagnetic Lagrangian}

The nematic liquid crystal is a uniaxial dielectric and diamagnetic anisotropic medium which determines a nontrivial structure of the constitutive tensor $\chi^{ijkl}$. The director $N^i$ determines the optical axis in the medium, so that the permittivity and permeability tensors are anisotropic: for matter at rest with the optical axis along $x$, 
\begin{equation}
\varepsilon^{ab} = \left(\begin{array}{ccc}\varepsilon_{\|} & 0 & 0\\
0 & \varepsilon_{\perp}& 0 \\ 0 & 0 &\varepsilon_{\perp}\end{array}\right),\qquad
\mu^{-1}_{ab} = \left(\begin{array}{ccc}\mu^{-1}_{\|} & 0 & 0\\
0 & \mu^{-1}_{\perp}& 0 \\ 0 & 0 &\mu^{-1}_{\perp}\end{array}\right),\label{epmuLQ}
\end{equation}
where the relative permittivity $\varepsilon_{\perp}$ in the plane perpendicular the optical axis is different from the relative permittivity $\varepsilon_{\|}$ in the direction along the optical axis, and the same is the case for the perpendicular and parallel relative permeability functions $\mu_{\perp}$ and $\mu_{\|}$. The dielectric and magnetic anisotropies are conveniently described by
\begin{align}
\Delta\varepsilon := \varepsilon_{\|}-\varepsilon_{\perp},\qquad
\Delta\mu^{-1} := \mu^{-1}_{\|}-\mu^{-1}_{\perp}.\label{delmu}
\end{align}
The latter notation should {\it not} be misunderstood as the inverse of the difference $\mu_{\|}-\mu_{\perp}$, that is, $\Delta\mu^{-1} \neq (\Delta\mu)^{-1}$. Strictly speaking, one should write (\ref{delmu}) as $\Delta(\mu^{-1})$, but we omit the parentheses to simplify the formulas. 

Like in the isotropic fluid, we assume that the permittivities $\varepsilon_{\|}(\nu), \varepsilon_{\perp}(\nu)$ and permeabilities $\mu_{\|}(\nu), \mu_{\perp}(\nu)$ may depend on the matter density $\nu$, giving rise to the possible eletro- and magneto-striction effects. 

The explicit form of the constitutive tensor, which takes into account all the necessary symmetries and which reproduces (\ref{epmuLQ}) in the rest frame of the medium, is given by
\begin{align}
\chi^{ijkl}={}& {\frac{1}{\mu_0\mu_{\|}}}\,(g^{ik}g^{jl}-g^{il}g^{jk}) + {\frac{1}{\mu_0c^2}}(n^2\mu^{-1}_{\perp}
- \mu^{-1}_{\|})\,(g^{ik} u^{j} u^{l}-g^{il} u^{j} u^{k} + g^{jl} u^{i} u^{k}-g^{jk} u^{i} u^{l})\nonumber\\
&+\,{\frac{1}{\mu_0}}\Delta\mu^{-1}\,(g^{ik}N^jN^l - g^{il}N^jN^k + g^{jl}N^iN^k - g^{jk}N^iN^l)\nonumber\\
&-\,{\frac{1}{\mu_0c^2}}\left(\Delta\varepsilon + \Delta\mu^{-1}\right)\,(u^iu^kN^jN^l - u^iu^lN^jN^k
+u^{j}u^{l}N^{i}N^{k}-u^{j}u^{k}N^{i}N^{l}).\label{explicit}
\end{align}
Here the refractive index $n^2 = \mu_\perp\varepsilon_\perp$. One can check that in the isotropic case, with $\varepsilon_{\|} = \varepsilon_{\perp} = \varepsilon$, $\mu_{\|} = \mu_{\perp} = \mu$, we recover (\ref{chi1}). 

With the help of the constitutive tensor (\ref{explicit}), we can use (\ref{CR}) to obtain an explicit expression for the electromagnetic excitation:
\begin{align}
H^{kl}={}&\frac{1}{\mu_0}\left(\mu^{-1}_{\perp} + \Delta\mu^{-1}\right)F^{kl}
+ \frac{2}{\mu_0}\Delta\mu^{-1}\,F^{[k}{}_n N^{l]}N^n\nonumber\\
{}&+ {\frac {2}{\mu_0c^2}}\left(\varepsilon_{\perp} - \mu^{-1}_{\perp} - \Delta\mu^{-1}\right)F^{[k}{}_n u^{l]}u^n
- {\frac {2}{\mu_0c^2}}\left(\Delta\varepsilon + \Delta\mu^{-1}\right)N^{[k}u^{l]}\,F_{pq}N^p u^q,\label{Hab}
\end{align}
and it is straightforward to write down the corresponding electromagnetic Lagrangian (\ref{Le}):
\begin{align}
{L}^{\rm e}{}=& -\,{\frac 1{4\mu_0}}(\mu^{-1}_{\perp}+\Delta\mu^{-1})F_{ij}F^{ij} - {\frac 1{2\mu_0}}
\Delta\mu^{-1}\,(F_{kl} N^l)^2\nonumber\\
{}&-\,{\frac 1{2\mu_0c^2}}\left(\varepsilon_{\perp} - \mu^{-1}_{\perp} - \Delta\mu^{-1}\right)(F_{kl}u^l)^2
+ {\frac 1{2\mu_0c^2}}\left(\Delta\varepsilon + \Delta\mu^{-1}\right)(F_{pq}N^p u^q)^2.\label{Lem1}
\end{align}
This Lagrangian looks especially clear when we rewrite it in terms of the electric and magnetic 4-vectors (\ref{EBvec4}):
\begin{align}
{L}^{\rm e}=\frac{1}{2}\left(\varepsilon_0\varepsilon^{ij}{\cal E}_{i}{\cal E}_{j}
- \mu_0^{-1}\mu^{-1}_{ij}{\cal B}^{i}{\cal B}^{j}\right),\label{4lem}
\end{align}
where $\varepsilon^{ij}$ is the 4-permittivity tensor and $\mu^{-1}_{ij}$ the inverse of the 4-permeability tensor, given by
\begin{align}
\varepsilon^{ij}:={}&-\varepsilon_{\perp}g^{ij}+\Delta\varepsilon N^{i}N^{j},\label{4epsilon}\\
\mu^{-1}_{ij}:={}&-\mu^{-1}_{\perp}g_{ij}+\Delta\mu^{-1} N_{i}N_{j}.\label{4mu}
\end{align}
The corresponding constitutive relation then reads, recall (\ref{DHvec4}):
\begin{equation}
{\mathcal D}^i = \varepsilon_0\varepsilon^{ij}{\mathcal E}_j,\qquad
{\mathcal H}_i = \mu_0^{-1}\mu^{-1}_{ij}{\mathcal B}^j.\label{CRLQ}
\end{equation}
In the isotropic limit ($\varepsilon_{\|} = \varepsilon_{\perp} = \varepsilon$, $\mu_{\|} = \mu_{\perp} = \mu$), one recovers (\ref{CRopt}).

We can now compute the variations of ${L}^{\rm e}$ in (\ref{Lem1}) with respect to the material variables:
\begin{align}\label{dLedu}
{\frac {\delta{L}^{\rm e}}{\delta u^i}} &= {\frac 1{\mu_0c^2}}\left(
\varepsilon_{\perp} - \mu^{-1}_{\perp} - \Delta\mu^{-1}\right)F_{ik}F^{kl}u_l - {\frac 1{\mu_0c^2}}
\left(\Delta\varepsilon + \Delta\mu^{-1}\right)(F_{pq} N^p u^q)\,F_{ik}N^k,\\
{\frac {\delta{L}^{\rm e}}{\delta N^i}} &= \frac{1}{\mu_0}\Delta\mu^{-1}\,F_{ik}F^{kl}N_l 
+{\frac 1{\mu_0c^2}}\left(\Delta\varepsilon +\Delta\mu^{-1}\right)(F_{pq}N^p u^q)\,F_{ik}u^k,\label{dLedN}\\
{\frac {\delta{L}^{\rm e}}{\delta \nu}} &= -\,{\frac 12}\left(\varepsilon_0\,{\frac {\partial\varepsilon}
{\partial\nu}}\,{\cal E}^2 + {\frac 1{\mu_0\mu_{\perp}^2}}\,{\frac {\partial\mu_{\perp}}{\partial\nu}}
\,{\cal B}^2\right) +  {\frac 12}\left(\varepsilon_0\,{\frac {\partial\Delta\varepsilon}
{\partial\nu}}\,({\cal E}_i N^i)^2 - {\frac 1{\mu_0}}\,{\frac {\partial\Delta\mu^{-1}}{\partial\nu}}
\,({\cal B}_i N^i)^2\right).\label{dLednu}
\end{align}

Now, using the expression (\ref{Hab}), we find explicitly the canonical energy-momentum tensor:
\begin{align}
{\stackrel{\rm M}\Sigma}{}_i{}^j={}&\frac{1}{\mu_0}\left(\mu^{-1}_{\perp} + \Delta\mu^{-1}\right)
\left[-\,F^{jk}F_{ik} + {\frac 14}\delta_i^j F^{kl}F_{kl}\right]\nonumber\\
&+\,{\frac {1}{\mu_0c^2}}\left(\varepsilon_{\perp} -\mu^{-1}_{\perp} - \Delta\mu^{-1}\right)\left[
-\,F^{jk}u_k F_{il}u^l + {\frac 12}\delta_i^j (F_{kl}u^l)^2 + u^j F_{ik}F^{kl}u_l\right]
\nonumber\\ &
+\frac{1}{\mu_0}\Delta\mu^{-1}\left[-\,F^{jk}N_k F_{il}N^l + {\frac 12}
\delta_i^j (F_{kl}N^l)^2 + N^j F_{ik}F^{kl}N_l\right]\nonumber\\
& +\,{\frac 1{\mu_0c^2}}\left(\Delta\varepsilon + \Delta\mu^{-1}\right)(F_{pq}N^p u^q)
\left[-\,{\frac 12}\delta_i^j (F_{kl}N^k u^l) - u^j F_{in}N^n + N^j F_{in}u^n\right],\label{TMLQ}
\end{align}
which is the Minkowski tensor of the electromagnetic field inside the liquid crystal medium.

Comparing (\ref{dLedu}) and (\ref{dLedN}) with (\ref{TMLQ}), we immediately
verify the correct balance equation for the electromagnetic angular momentum part of the system,
\begin{align}
{\stackrel{\rm M}\Sigma}{}_{[ij]} = {\frac {\delta{L}^{\rm e}}
{\delta u^{[i}}}\,u_{j]} + {\frac {\delta{L}^{\rm e}}{\delta N^{[i}}}\,N_{j]}.\label{angularM}
\end{align}
This is in perfect agreement with the Noether theorem.

Substituting (\ref{Hab}) into (\ref{TA}), and making use of (\ref{dLedu})-(\ref{TMLQ}), we derive for the Abraham energy-momentum tensor:
\begin{equation}
{\stackrel{\rm A}\Sigma}{}_k{}^i = {\stackrel{\rm M}\Sigma}{}_k{}^i - \varepsilon_0\Delta\varepsilon
({\mathcal E}_jN^j){\mathcal E}_{[k}N^{i]} + \mu_0^{-1}\Delta\mu^{-1}({\mathcal B}_jN^j){\mathcal B}_{[k}N^{i]} 
- u^i\biggl[P_k^j{\frac {\delta{L}^{\rm e}}{\delta u^j}} - N_k{\frac {u^j}{c^2}}\,
{\frac {\delta{L}^{\rm e}}{\delta N^j}}\biggr].\label{TAMQ}
\end{equation}
As a result of the field equations (\ref{EqNN}), we then compute the total canonical energy-momentum tensor of the coupled system ``matter+field'':
\begin{align}
\sigma_k{}^i + {\stackrel{\rm M}\Sigma}{}_k{}^i = {\frac {\rho}{c^2}}u_ku^i - p^{\rm eff}\left(\delta_k^i
- {\frac {u^iu_k}{c^2}}\right) + {\stackrel{\rm s}\sigma}{}_k{}^i - P^j_{[k}{\frac {\delta{L}^{\rm m}}
{\delta N^j}}N^{i]} + {\stackrel{\rm A}\Sigma}{}_k{}^i,\label{totLQ}
\end{align}
where with the help of (\ref{dLednu}) we find an explicit effective pressure (\ref{peff})
\begin{equation}\label{peffLQ}
p^{\rm eff} = p+\,{\frac 12}\nu\left(\varepsilon_0\,{\frac {\partial\varepsilon}{\partial\nu}}\,{\cal E}^2
+ {\frac 1{\mu_0\mu_{\perp}^2}}\,{\frac {\partial\mu_{\perp}}{\partial\nu}}\,{\cal B}^2\right) - {\frac 12}
\nu\left(\varepsilon_0\,{\frac {\partial\Delta\varepsilon}{\partial\nu}}\,({\cal E}_i N^i)^2
- {\frac 1{\mu_0}}\,{\frac {\partial\Delta\mu^{-1}}{\partial\nu}}\,({\cal B}_i N^i)^2\right),
\end{equation}
that accounts for the electro- and magnetostriction effects, cf. (\ref{presseff}). 

Similarly to the spinning fluid case (\ref{Ttot}) and (\ref{kinS}), we can introduce the {\it kinetic energy-momentum tensor} of the liquid crystal medium 
\begin{equation}\label{kinLQ}
\kappa_k{}^i := {\frac \rho {c^2}}u_ku^i - p^{\rm eff}\left(\delta_k^i - {\frac {u_ku^i}{c^2}}\right)
+ {\stackrel {\rm s}\sigma}{}_k{}^i - P^j_{[k}{\frac {\delta{L}^{\rm m}}{\delta N^j}}N^{i]},
\end{equation}
so that the right-hand side of (\ref{totLQ}) becomes the sum of the kinetic and Abraham tensors $\kappa_k{}^i + {\stackrel{\rm A}\Sigma}{}_k{}^i$. Just like (\ref{kinS}), the energy-momentum (\ref{kinLQ}) is not symmetric due to a nontrivial spin of the liquid crystal (\ref{spin3}), however the angular momentum of matter is conserved 
\begin{align}
\kappa_{[ij]} + \partial_k S_{ij}{}^k = 0.\label{kinAL}
\end{align}

\section{General model of closed system of interacting matter and electromagnetic field}\label{general}

In Sec. \ref{weyss} and Sec. \ref{crystal} we have analysed the spinning fluid and the liquid crystal medium as the two special examples of the complex matter. The results obtained can be further extended to the general case of a closed system of complex medium coupled to the electromagnetic field.

\subsection{Relativistic Lagrangian for general linear medium}

When dealing with the electromagnetic field as an {\it open system}, the properties of light propagating in a material medium can be consistently understood by treating the components of the constitutive tensor $\chi^{ijkl} = \chi^{ijkl}(t,\bm{x})$ as a {\it background external field} in the electromagnetic Lagrangian (\ref{Le}). 

To deal with an arbitrary {\it closed system} of interacting matter and field, we do not view the constitutive tensor as a dynamical variable by itself, but rather consider it as a function of the fundamental matter fields 
\begin{align}
\chi^{ijkl} = \chi^{ijkl}(u^i,\nu,\Psi^A).\label{chi}
\end{align}
Quite generally, we assume that the constitutive tensor depends on the set of {\it material variables}: the matter velocity field $u^{i}$, the particle number density $\nu$, and some additional fields $\Psi^A(t, \bm{x})$ that describe the {\it internal degrees of freedom} of a medium. For example, in the spinning fluid model $\Psi^A = \{\mu^{AB}, b^i_A\}$, whereas for the liquid crystal $\Psi^A = \{N^i\}$ (and in both cases, one should also include here the Lagrange multipliers that impose the appropriate constraints). The total Lagrangian of the closed system of interacting matter and the electromagnetic field reads
\begin{align}
L = {L}^{\rm m} + {L}^{\rm e},\label{Ltot}
\end{align}
where the electromagnetic field Lagrangian is given by (\ref{Le}) with the constitutive tensor (\ref{chi}), whereas the matter Lagrangian is a sum
\begin{eqnarray}\label{Lmat}
{L}^{\rm m} = -\,\rho(\nu,s) + {L}^{\rm ani}(\nu,\partial_i\nu,u^i,\partial_ju^i,\Psi^A,\partial_i\Psi^A).
\end{eqnarray}
of the {\it internal energy density} $\rho(\nu,s)$ of an ideal fluid, and the second  term ${L}^{\rm ani}$ accounting for the intrinsic dynamics of the medium responsible for its anisotropic properties. For the particular examples, see (\ref{L}) and (\ref{Lfluid}) above. 

The dynamics of the closed ``matter+field'' system, described by the set of variables $\Phi^A = (A_i, u^i, \nu, s, \Psi^A)$, is then determined by the Euler-Lagrange equations of motion $\delta{L}/\delta \Phi^A = 0$. The standard Lagrange-Noether analysis of the general model (\ref{Lmat}) gives rise to the canonical energy-momentum tensor of matter \cite{JOPT}
\begin{align}\label{Tmat}
\sigma{}_k{}^i = u^i{\cal P}_k - p^{\rm eff}\left(\delta_k^i - {\frac {u_ku^i}{c^2}}\right) + {\stackrel{\rm a}{\sigma}}{}_k{}^i,
\end{align}
where (speaking qualitatively) the structure of the elastic stress ${\stackrel{\rm a}{\sigma}} {}_k{}^i = {\stackrel{\rm a}{\sigma}}{}_k{}^i(\Psi)$ is determined by the $\Psi$ fields and the Lagrangian $L^{\rm ani}$, the 4-momentum ${\cal P}_k = {\frac 1{c^2}}\rho u_k\,+ $ {\it electromagnetic} field contribution, and the effective pressure $p^{\rm eff} = p\,+$ electric and magnetic {\it striction} terms. 

On the other hand, from first principles in the framework of the Lagrange-Noether formalism we find the Minkowski tensor (\ref{TM}) as the canonical energy-momentum current of the electromagnetic field. In general, both canonical tensors of matter $\sigma{}_k{}^i$ and of the electromagnetic field ${\stackrel{\rm M}\Sigma}_k{}^i$ (=Minkowski tensor) are not symmetric, when the material spin density $S_{ij}{}^k$ is nontrivial, so that the {\it total energy-momentum} tensor $T_k{}^i = \sigma_k{}^i + {\stackrel{\rm M}\Sigma}{}_k{}^i$ is also not symmetric. However, the total energy-momentum current and the total angular momentum are perfectly conserved:
\begin{equation}
\partial_jT_i{}^j = 0,\qquad T_{[ij]} + \partial_k S_{ij}{}^k = 0.\label{Ttot0}
\end{equation}
No forces and torques appear on the right-hand sides, because the ``matter+field'' system is {\it closed}.

\subsection{Kinetic tensor: key to Abraham}

The Minkowski energy-momentum tensor arises from first principles as a Noether canonical current. What can be said of the Abraham tensor -- how physically relevant it is? Can one recover it in a certain natural way?

Let us define the {\it kinetic material energy-momentum} tensor:
\begin{align}
\kappa_k{}^i &= u^i{\stackrel{\rm kin}{\cal P}}{\!}_k - p^{\rm eff}\left(\delta_k^i - {\frac {u_ku^i}{c^2}}\right)
+ \stackrel{\rm a}{\sigma}_k{\!}^i - {\frac{\delta{L}^{\rm ani}}{\delta \Psi^A}}(\ell_k{}^i)^A_B\Psi^B,\\
{\stackrel{\rm kin}{\cal P}}{\!}_i &= \frac{\rho}{c^2} u_i + {\frac{u^k} {c^2}}
{\frac{\delta{L}^{\rm ani}}{\delta \Psi^A}}(\ell_{ik})^A_B\Psi^B.
\end{align}
Here $(\ell_{ik})^A_B$ are the generators of the Lorentz algebra in the corresponding (reducible) representation of the matter fields $\Psi^A$.

The explanation of the physical relevance of the Abraham energy-momentum tensor is as follows \cite{JOPT}: The total energy-momentum tensor of the closed ``matter+field'' system
\begin{align}
T_k{}^i = {\stackrel {\rm M}\Sigma}_k{}^i + \sigma_k{}^i  
= {\stackrel {\rm A}\Sigma}_k{}^i + \kappa_k{}^i\label{Tdec}
\end{align} 
admits two alternative decompositions either into a sum of the canonical energy-momentum tensor of matter plus the Minkowski tensor (``Minkowski decomposition''), or into a sum of the kinetic energy-momentum tensor of matter plus the Abraham tensor (``Abraham decomposition''). Both decompositions yield the correct total energy-momentum tensor of the closed system. For {\it simple media} without microstructure we find the remarkable structure \cite{obukhov1}, see also \cite{Partanen:2019},
\begin{equation}
\kappa_k{}^i = u^i{\stackrel{\rm kin}{\cal P}}{\!}_k - p^{\rm eff}\left(\delta_k^i - {\frac {u_ku^i}{c^2}}\right),
\qquad {\stackrel{\rm kin}{\cal P}}{\!}_k = \frac{\rho}{c^2} u_k,
\end{equation}
where the first term on the right-hand side reproduces the usual kinetic ``mass$\times$velocity'' momentum.

\subsection{Resolving Minkowski-Abraham controversy}

Comparing the two energy-momentum tensors, we observe a solid status of the Minkowski tensor as a canonical Noether current derived {\it from first principles}. In contrast, the Abraham tensor is (a) {\it ad hoc} construct, (b) {\it derived} from the Minkowski tensor. 

Nevertheless, both objects are physically relevant if one carefully distinguishes the two physical situations for open and closed systems.

(I) When the matter is a {\it non-dynamical background}, the electromagnetic field is consistently treated as an {\it open system} with the constitutive tensor $\chi^{ijkl}$ as a fixed external field. Then the {\it Minkowski tensor} gives the correct energy and momentum for the light in matter.
 
(II) A {\it dynamical matter} (especially, a moving one) forms a {\it closed system} together with the coupled electromagnetic field. Then only the {\it total energy-momentum} tensor has the physical meaning. 

Finally, {\it whence Abraham?} In the decomposition of the total energy-momentum (\ref{Tdec}), the Abraham tensor precisely absorbs all the electromagnetic terms, whereas the rest of $T_k{}^i$ turns out to be the {\it kinetic} energy-momentum tensor $\kappa_k{}^i$ that depends on matter only.

\section{Conclusions and outlook}\label{conc}

One can develop a consistent relativistic Lagrangian model for the general linear complex medium coupled to the electromagnetic field on the basis of the covariant approach \cite{Birkbook,JOPT}. The experiments with light in matter can be correctly analysed by using the Minkowski momentum when the wave field is treated as an open system on a fixed matter background. For the case of a closed system, when both matter and electromagnetic field are dynamical, one can use any of the energy-momentum tensors -- both Minkowski and Abraham momenta provide equally correct theoretical explanations. 

The {\it kinetic} material energy-momentum tensor $\kappa_k{}^i$ is a convenient object that underlies the fundamental decomposition (\ref{Tdec}). Summarizing, we conclude that both Minkowski and Abraham tensor can be consistently used, provided one carefully addresses the dynamics of matter. It is impossible to select a unique ``correct'' electromagnetic momentum: the Minkowski tensor ${\stackrel {\rm M}\Sigma}{}_k{}^i$ is the fundamental canonical current, whereas Abraham's ${\stackrel {\rm A}\Sigma}{}_k{}^i$ is a useful effective construct -- a ``purely electromagnetic'' piece of total energy-momentum tensor, complemented by a ``purely kinetic'' $\kappa_k{}^i$ of the medium. 

It is important to stress that the symmetry of the energy-momentum tensor is {\it not} a fundamental property. In this sense, the main argument which often underlies the choice of the Abraham tensor is invalid, in general, and at most it applies to {\it simple} media. As soon as we deal with a {\it complex} matter that has microstructure, the canonical tensor of spin $S_{ij}{}^k$ is nontrivial, and hence the canonical energy-momentum tensor is not symmetric, $\sigma_{[ij]} + \partial_k S_{ij}{}^k = 0$, and the same of course applies to the total energy-momentum tensor of the closed system (\ref{Ttot0}). However, one can {\it relocalize} the canonical tensor of energy-momentum and spin \cite{hehl76}, and a particular relocalization of Belinfante and Rosenfeld \cite{Belinfante1,Belinfante2,Rosenfeld} introduces the symmetrized energy-momentum
\begin{equation}\label{BR}
{\stackrel {\rm BR}\sigma}{}_k{}^i = \sigma_k{}^i -\partial_j\left(S^{ij}{}_k + S_k{}^{ji} - S_k{}^{ij}\right).
\end{equation}
By means of this procedure, the total energy-momentum (\ref{Tdec}) is replaced by the corresponding symmetric and conserved energy-momentum tensor
\begin{equation}
\partial_j{\stackrel {\rm BR}T}{}_i{}^j = 0,\qquad {\stackrel {\rm BR}T}{}_{[ij]} = 0.\label{TtotBR}
\end{equation}
The construction of the symmetric kinetic energy momentum tensor $\kappa_k{}^i$ via the Belinfante-Rosenfeld relocalization is discussed in detail in \cite{JOPT}.

In this relation, a remark is in order about the extension of the special-relativistic formalism, which we considered here, to the general-relativistic framework in which the spacetime metric is no longer fixed $g_{ij} = {\rm diag}(c^2, -1, -1, -1)$ but becomes a dynamical variable $g_{ij}(t, {\bm x})$ that describes the gravitational field. Quite remarkably, the Abraham energy-momentum tensor can be recovered then from the variational derivative of the matter Lagrangian with respect to the metric, see \cite{antoci1,antoci2,antoci3} for the case of simple media and \cite{Dereli1,Dereli2,ann2012} for the complex media.

For the sake of clarity, our discussion was confined here to the media without dissipation and dispersion. The corresponding analysis of electrodynamics in dispersive matter can be found in \cite{Furutsu,Dewar,garrison,philbin,philbin2,Bliokh:2014,Bliokh:2017,bradshawboydmilonni,campbell,shevchenko2,Partanen:2021}, mostly for the simple media though, and it would interesting to generalize the treatment to the case of arbitrarily moving complex media.

\end{document}